\documentclass[twocolumn]{aastex62} 
\usepackage{amsmath,amstext}
\usepackage{stackengine}
 \usepackage{mathtools}
\usepackage{graphicx}
\usepackage{txfonts}

\newcommand{\noop}[1]{}

\newcommand{\be}{\begin{eqnarray}}
\newcommand{\ee}{\end{eqnarray}}

\newcommand{\MSun} {\mbox{$M_{\odot}$}}

\shorttitle{ Towards a disc lifetime distribution}
\begin{document}

\title{Low-mass stars: Their Protoplanetary Disc Lifetime Distribution }

\author[0000-0002-5003-4714]{Susanne Pfalzner} 
\affiliation{J\"ulich Supercomputing Center, Forschungszentrum J\"ulich, 52428 J\"ulich, Germany}
\affiliation{Max-Planck-Institut f\"ur Radioastronomie, Auf dem H\"ugel 69, 53121 Bonn, Germany}
\author{Furkan Dincer} 
\affiliation{J\"ulich Supercomputing Center, Forschungszentrum J\"ulich, 52428 J\"ulich, Germany}
\affiliation{RWTH Aachen, Aachen, Germany}

\email{s.pfalzner@fz-juelich.de}


\begin{abstract}

While most protoplanetary discs lose their gas within less than 10 Myr, individual disc lifetimes vary from \mbox{$<$ 1 Myr} to $\gg$ 20 Myr, with some discs existing for \> 40 Myr.   Mean disc half lifetimes hide this diversity; only a so-far non-existing disc lifetime distribution could capture this fact. The benefit of a disc lifetime distribution would be twofold. First, it provides a stringent test on disc evolution theories. Second, it can function as input for planet formation models. Here, we derive such a disc lifetime distribution. We heuristically test different standard distribution forms for their ability to account for the observed disc fractions at certain ages. Here, we concentrate on the distribution for low-mass stars (spectral type M3.7--M6, $M_s \approx $ 0.1 -- 0.24 \MSun) because disc lifetimes depend on stellar mass. A Weibull-type distribution ($k$=1.78, $\lambda$=9.15) describes the observational data if all stars have a disc at a cluster age $t_c$=0. However,  a better match exists for lower initial disc fractions. For f(t=0)= 0.65, a Weibull distribution ($k$=2.34, $\lambda$=11.22) and a Gauss distribution ($\sigma$=9.52, $\mu$=9.52) fit similarly well the data. All distributions have in common that they are wide, and most discs are dissipated at ages \mbox{$>$ 5 Myr.} The next challenge is to quantitatively link the diversity of disc lifetimes to the diversity in planets.
\end{abstract}

\keywords{circumstellar matter, protoplanetary discs, open clusters and associations, planet formation}

\section{Introduction}
\label{sec:intro}

Planets form from the gas and dust discs surrounding young stars. After some time, the gas in the discs dissipates, basically stopping the planets from accreting additional gas. Therefore, the length of the period available for the planets to build up is a crucial parameter for planet formation theory. 

There is ample observational evidence that discs develop over time. For example, the mean millimetre dust mass of discs decreases with the age of the star-forming region \citep[e.g., ][]{Ansdell:2017, Andrews:2020}. Eventually, processes like photo-evaporation, winds, and dust growth  \citep[e.g.][]{Williams:2011, Ercolano:2017, Kunitomo:2020} lead to complete disc dissipation. Consequently, the disc fraction in star-forming regions decreases with cluster age. This decline is visible for diverse disc indicators, such as infrared excess or accretion signatures \citep{Haisch:2001, Hernandez:2007, Fedele:2010, Ribas:2014, Richert:2018}.

 \citet[][]{Haisch:2001} introduced determining the disc lifetime from the decline of the disc fraction with cluster age. While they used a linear fit to their data, exponential fits became the norm when more data became available. Different studies differ in the resulting median disc lifetime from  \mbox{$f_d$ = 1 -- 3 Myr} \citep[][]{Haisch:2001, Hernandez:2008,Fedele:2010,Ribas:2014,Richert:2018,Briceno:2019} to $f_d$ = 5 -- 10 Myr \citep[][]{Pfalzner:2014,Ribas:2015,Michel:2021,Pfalzner:2022}. 
 The differences mainly result from the different selection criteria for the included clusters and the different methods and evolutionary models used to estimate the ages of the clusters.

While determining a median disc lifetime demonstrated that planets form relatively fast, its limitations start to show. Several relatively old discs ($>$ 10 Myr) have recently been discovered  \citep[for example, ][]{Silverberg:2020}. Meanwhile, fully formed planets have been reported in much younger systems. The co-existence of short- and long-lived discs strongly indicates that the distribution of disc lifetimes likely is very wide. Thus, the notion of a median disc lifetime has limited usage power. Here, we demonstrate a method for deriving a disc lifetime distribution from existing disc fraction values. Such a disc lifetime distribution is essential for deepening our understanding of planet formation for two reasons: First, the disc lifetime distribution puts stringent constraints on disc dispersal models. Second, it can shed new light on the diversity of the planet formation processes \citep{{Mordasini:2009, Ida:2010, Mordasini:2012, Forgan:2013, Schlecker:2022, Emsenhuber:2023}}.

\section{Old discs and young planets}
\label{sec:old_discs}

Examples of discs older than 10 Myr are the \mbox{$\sim$ 10 Myr-old} \mbox{HD 98800 B} \citep{Ronco:2021}, the $\sim$ 14-Myr-old star \mbox{HD 139614} and HD 143006  \citep{Kennedy:2019,Ballabio:2021}. 
Moreover, a few $\sim$ 50 Myr-old stars exhibit large fractional IR luminosity indicative of a circumstellar disc and simultaneously spectroscopic signs of accretion. Examples are WISE J080822.18-644357.3 in the Carina association \mbox{($\approx$ 45 Myr)} \citep[][]{Silverberg:2016} \citep[][]{Murphy:2018} and the $\approx$ 55 Myr-old 2MASS J15460752-6258042 \citep[][]{Lee:2020}.

Examples of old discs always face the problem of distinguishing between debris and protoplanetary discs. Such a distinction is complicated by dust in protoplanetary discs being not primordial but subject to a continual cycle of growth and destruction \citep[][]{Wyatt:2015}. Nevertheless, many relatively old stars have been confidently identified as being surrounded by protoplanetary discs. \citet[][]{Silverberg:2020} and \citet[][]{Lee:2020} discovered six more accreting stars of ages $\approx$ 50 Myr. These objects likely represent long-lived CO-poor primordial discs or "hybrid" discs simultaneously exhibiting debris and primordial-disc features. All these very long-lived discs have in common that their host is typically an M-type star.

ALMA observations reveal that many discs display ring structures. A popular interpretation is that these ring structures could be caused by planets that have already grown to a substantial size or are pressure bumps \citep[for a summary, see][]{Marel:2023}. If ring structures are indeed an indicator of an advanced stage of planet formation, then they are also pointing at a wide spread in planet formation time scales. The most prominent two examples of ring-structured discs -- HL Tau  ($<$ 1 Myr) and TW Hydrae \mbox{($\sim$ 10 Myr)} -- differ considerably in their ages. 

The term  "the disc lifetime", when referring to the median disc lifetime, implicitly implies that discs exist for a typical characteristic period. The above examples make it clear that this is, to some extent, misleading. These detections of protoplanetary discs older than 20 Myr show that a shift towards a description by a disc lifetime distribution is long overdue. Such a disc lifetime distribution quantifies the relative occurrence rate of any disc lifetime, including those of very short and long-lived discs. 

Disc fractions in clusters are known to depend on stellar mass \citep[e.g.,][]{Carpenter:2006,Roccatagliata:2011,Yasui:2014,Ribas:2015,Richert:2018}. In a given cluster, high-mass stars tend to have considerably lower disc fractions than low-mass stars. Here, we concentrate on low-mass stars to separate the stellar mass dependence from the intrinsic spread in disc lifetimes. Looking only at low-mass stars (spectral type \mbox{M3.7 -- M6,} 
\mbox{$M_s \approx$ 0.1 -- 0.24 \MSun\ },
\citet[][and references therein]{Luhman:2023b}. We will see that the extensive spread in disc lifetimes persists. We chose this particular mass bin because it shows the highest disc fraction at all ages and is still 90\% complete for the investigated clusters (see section 5.1).

\section{Towards a disc lifetime distribution} 

The starting point for determining the disc lifetime distribution is the disc fractions in clusters of different ages. The disc fraction,  $f_d(t)$,  declines by the number of discs that have reached the end of their lifetime as
\be
f_d(t) = f_d(t=0) - \int_0^t T_d dt \label{eq:fractions},
\ee
where  $T_d$ denotes the individual disc lifetime and $f_d(0)$ the initial disc fraction, which might be 100\% or less.  If each star is surrounded by a disc at cluster age ($t$=0), this simplifies to $ f_d(t) = 1 - \int_0^t T_d(t) dt $. The disc lifetime distribution is
\be T_d(t) =  - \frac{d}{dt} f_d(t). \label{eq:dist} \ee
The problems in determining $T_d(t)$ are (1) the uncertainties in cluster ages and disc fractions resulting in a considerable scatter among the data points and (2) the under-representation of clusters older than 4 Myr. 
Unfortunately, linear and exponential fits to the decline in disc fraction with cluster age lead to counter-intuitive disc lifetime distributions. 

\begin{figure}[t]
\includegraphics[width=0.48\textwidth]{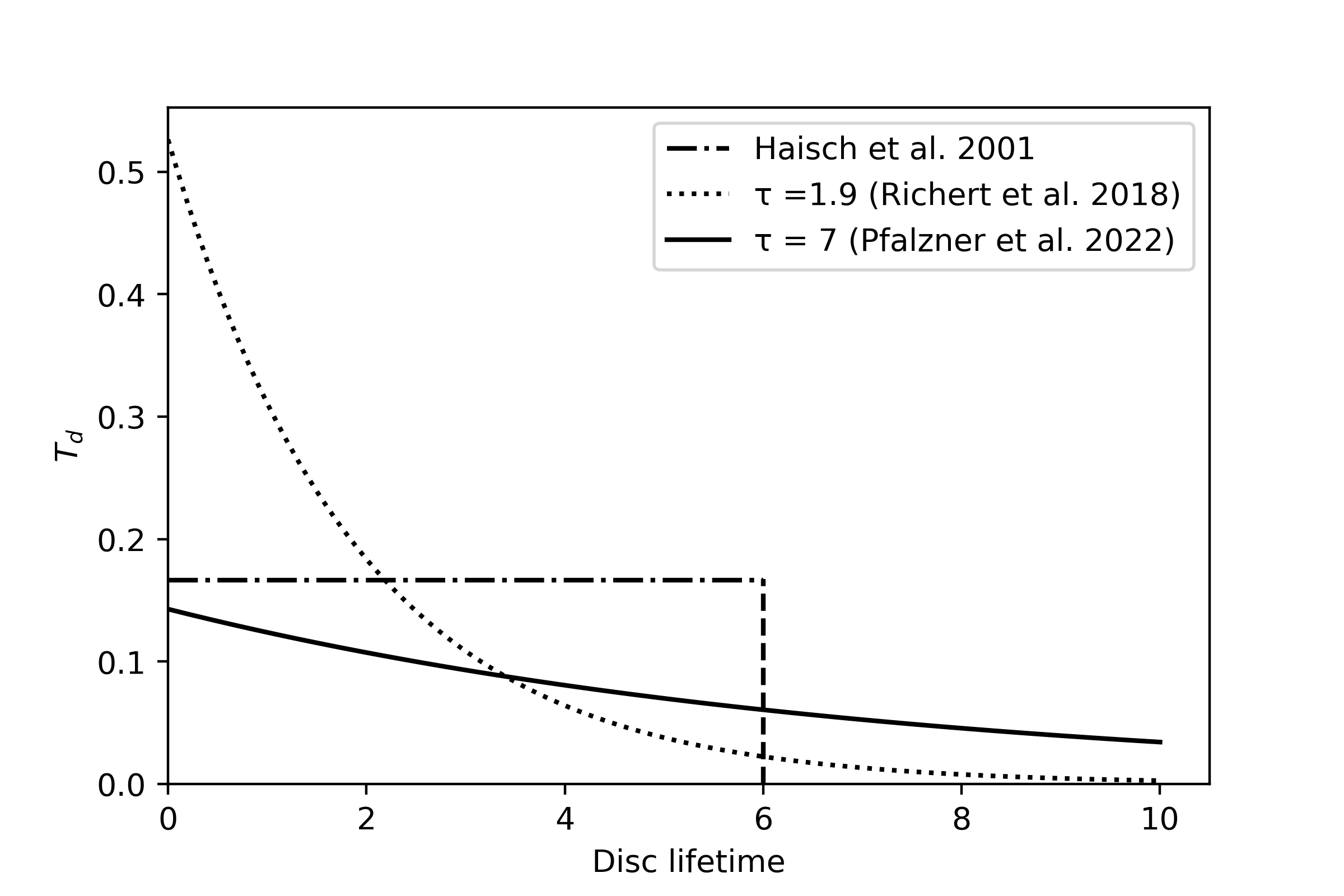}
\includegraphics[width=0.48\textwidth]{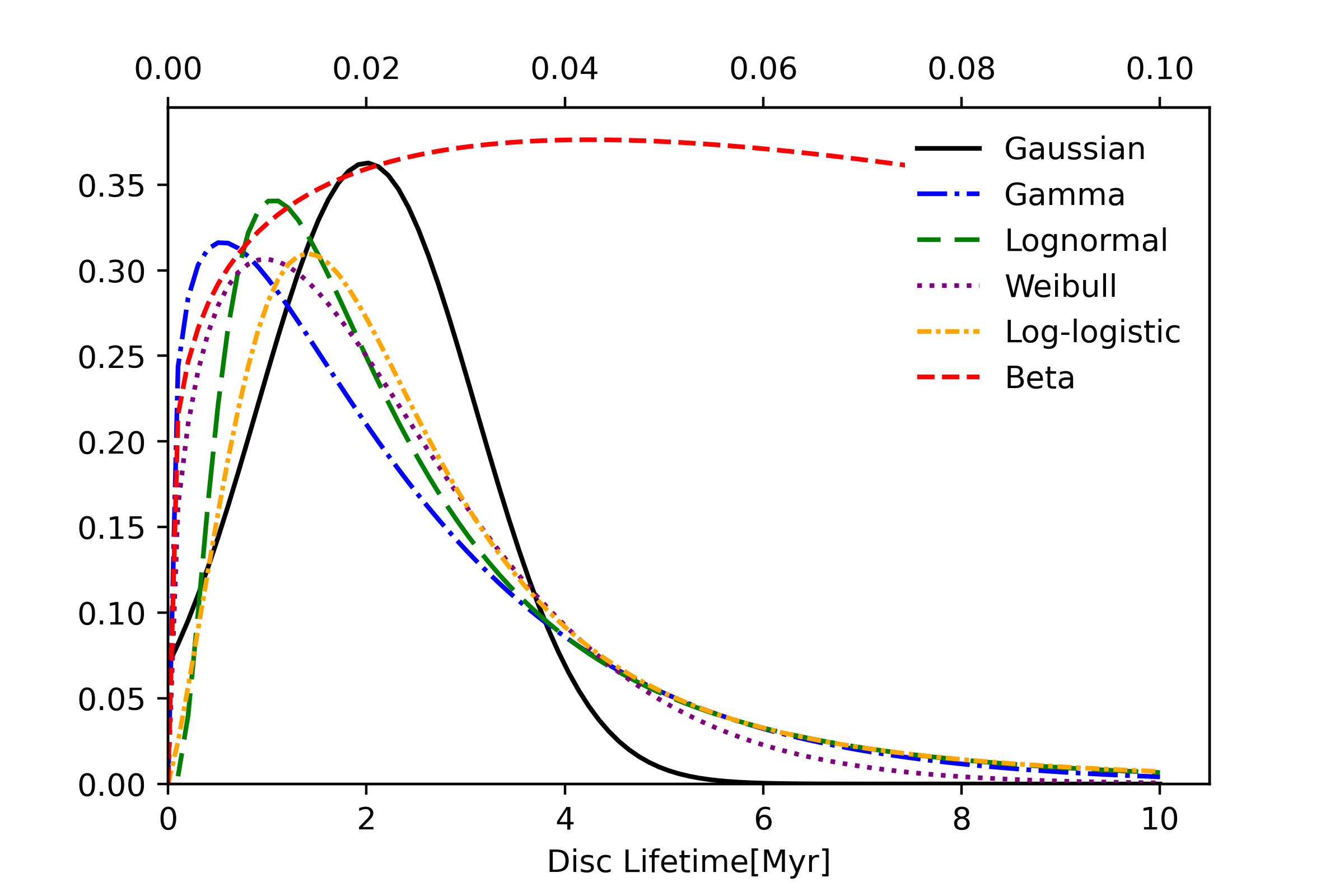}
\includegraphics[width=0.48\textwidth]{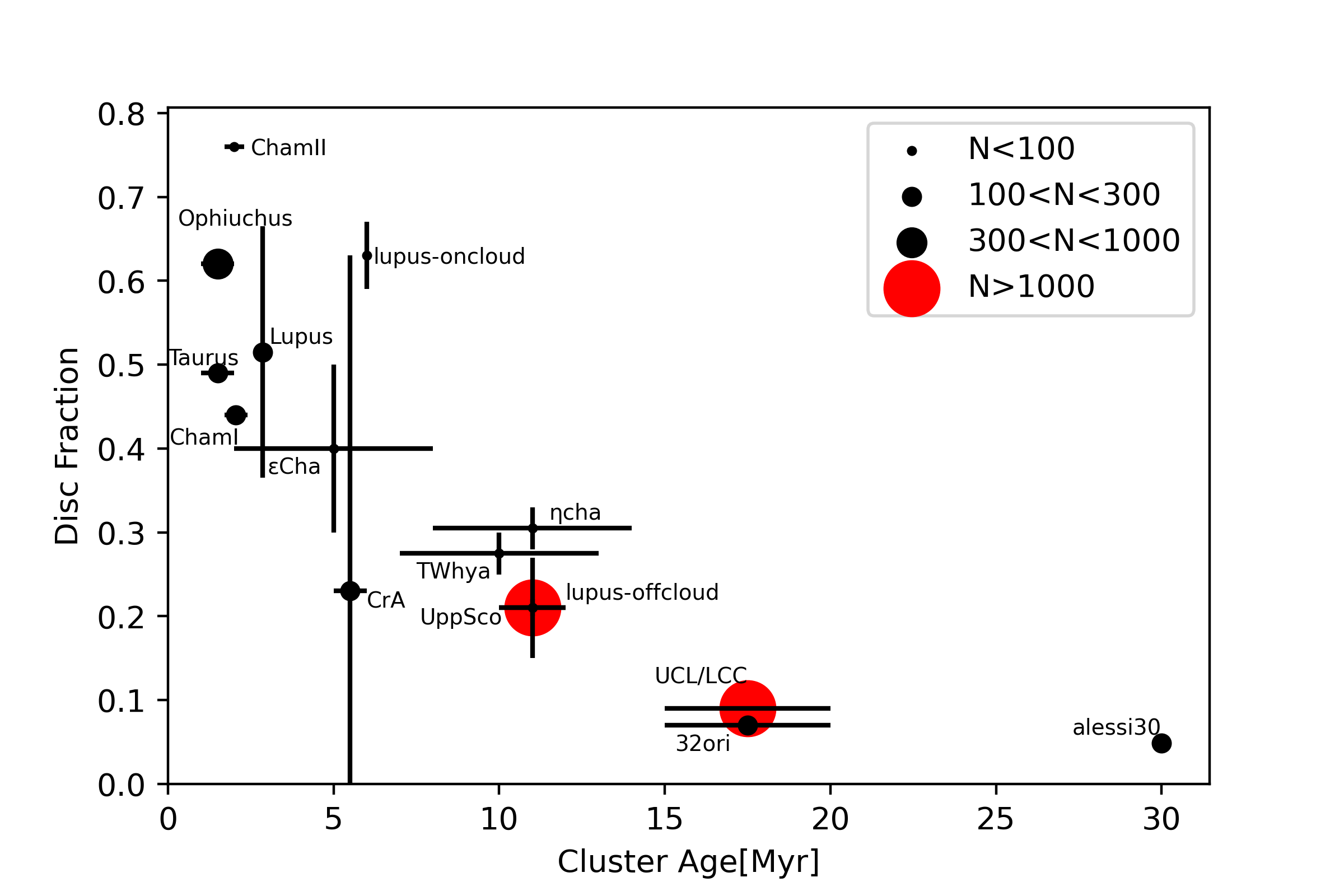}
\caption{Top: Disc lifetime distributions corresponding to the linear fits of the disc fraction vs age by \citet[][]{Haisch:2001} (black) and the exponential fits applied by \citet[][]{Richert:2018} (blue) and \citet[][]{Pfalzner:2022} (red). Middle: Schematic illustration of the distributions tested here. Bottom: Illustration of the data points' relative significance due to the cluster's number of stars. Only clusters within a distance, $d <$ 200 pc from the Sun, are shown. } 
\label{fig:distribution_schematic}
\end{figure}

\citet[][]{Haisch:2001} connected an assumed 100 \% initial disc fraction to the time when all discs seemed to have vanished (6 Myr) by a linear fit. The corresponding disc lifetime distribution is 
$T_d(t) = 1/6$ 
(see Fig. \ref{fig:distribution_schematic} top.) This constant distribution would imply that a disc is as likely to disperse at an age of 0.01 Myr as at 6 Myr. Such independence on age is conceptually unsatisfying. The idea of a typical disc lifetime reflects the expectation of a distribution that shows a clear maximum. A constant disc mass distribution fails to provide such a maximum in the distribution.  

As more cluster disc fractions became available, it became apparent that a linear fit was unsuitable. The question is whether the afterwards used exponential fits for the disc fraction lead to a more realistic disc lifetime distribution. In this case, the disc lifetime distribution would also be an exponential function of the form:
\be T_d(t) = \tau \exp(-t/\tau).\ee
It implies that discs are more likely to be dispersed when they are very young ($<$ 0.5 Myr) than when they are older (see Fig. \ref{fig:distribution_schematic} top). For example, the exponential fit given by \citet{Richert:2018} would imply that about 2.5 times more discs disperse when they are $<$ 0.5 Myr old than when they are  
\mbox{2 -- 2.5 Myr} old. For the exponential given in \citep{Pfalzner:2022}, the dominance of very early disc destruction is less severe; however, like any exponential fit, it also suffers from the problem that very young discs are the ones most likely to be dispersed, and there exists no maximum in the distribution being representative for a typical disc lifetime.

\begin{table}[t]
        \caption{Parameters for Gaussian fits.}
        \centering\begin{tabular}{lrcllccclccccc}
        \hline
        ID & $f_{USco}$ & $f_{U/L}$ & $t_{USco}$ & $t_{U/L} $& $\bar{t_d}$ & $\sigma(t_d)$ \\ 
        \hline
        L1$_{100}$  & 0.25 & 0.09  & 11.0 & 15.0 &  6.95  & 6.0 \\
        L2$_{100}$  & 0.22 & 0.09  & 11.0 & 17.0 &  2.85  & 10.55 \\
        L3$_{100}$  & 0.25 & 0.09  & 12.0 & 15.0 &  8.96  & 4.5 \\
        L4$_{100}$  & 0.22 & 0.11  & 11.0 & 15.0 &  4.20  & 8.8 \\
        L5$_{100}$  & 0.22 & 0.09  & 11.0 & 15.0 &  5.7   & 7.0  \\
        L1++ & 0.25 & 0.09  & 11.0 & 15.0 &  6.16 &  6.78
         \\
 \hline      
              \end{tabular}
\begin{flushleft}              
\tablecomments{
         Here Col.~1 indicates the model identifier, $f_{USco}$  stands for the assumed disc fraction of Upper Sco, $f_{U/L}$  for the disc fraction of UCL/LCC,  $t_{USco}$ the cluster age of Upper Sco, $t_{U/L} $ that of UCL/LLC, $\bar{t_d}$, the median age of the best-fit distribution and $\sigma(t_d)$ its standard deviation. L1++ is the best fit when the
clusters in \citet{Pfalzner:2022} with $d <$ 200 pc are considered in addition.}
        \label{tab:fit_gauss}
 \end{flushleft}        
\end{table}

\begin{figure}[t]
\includegraphics[width=0.48\textwidth]{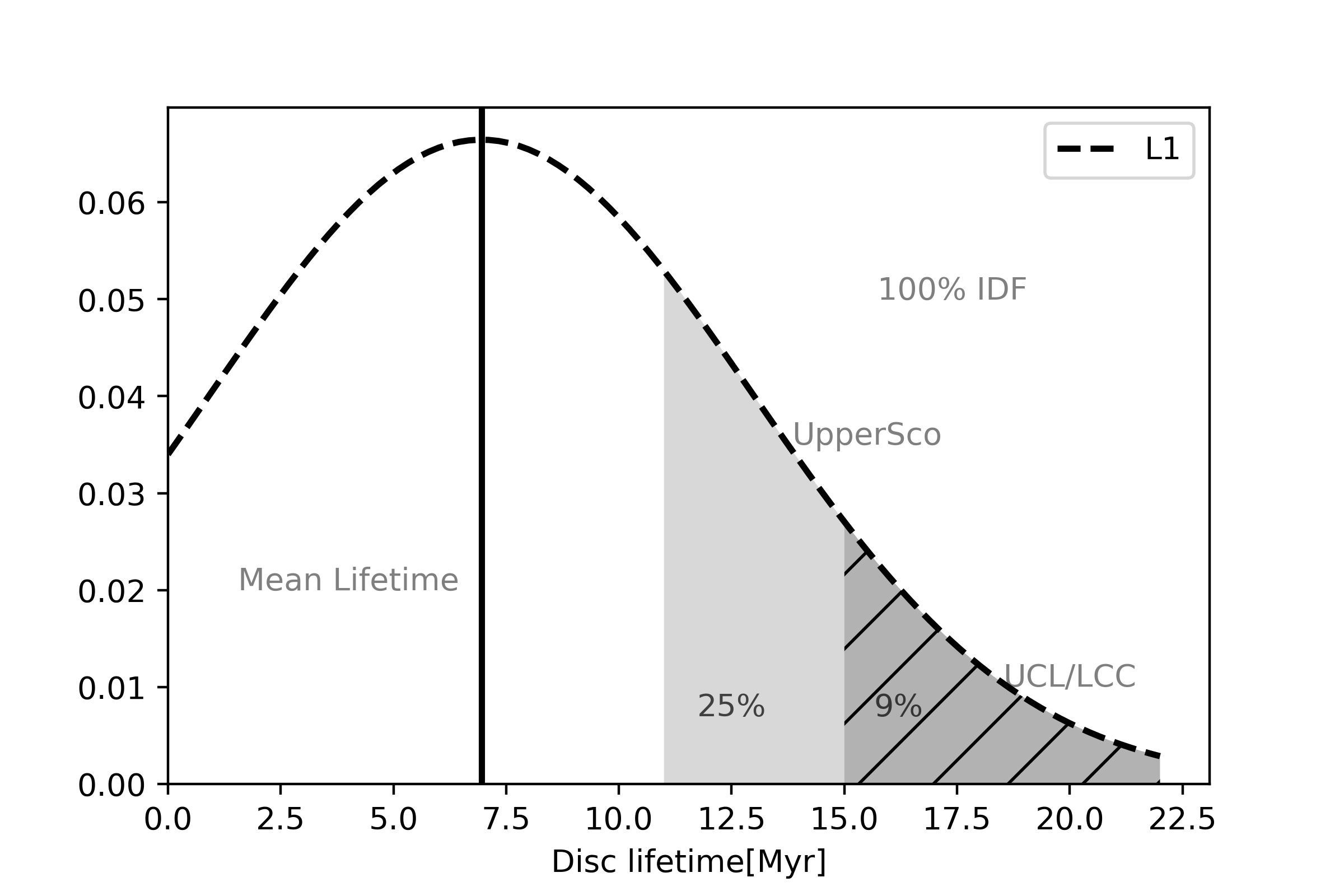}
\includegraphics[width=0.48\textwidth]{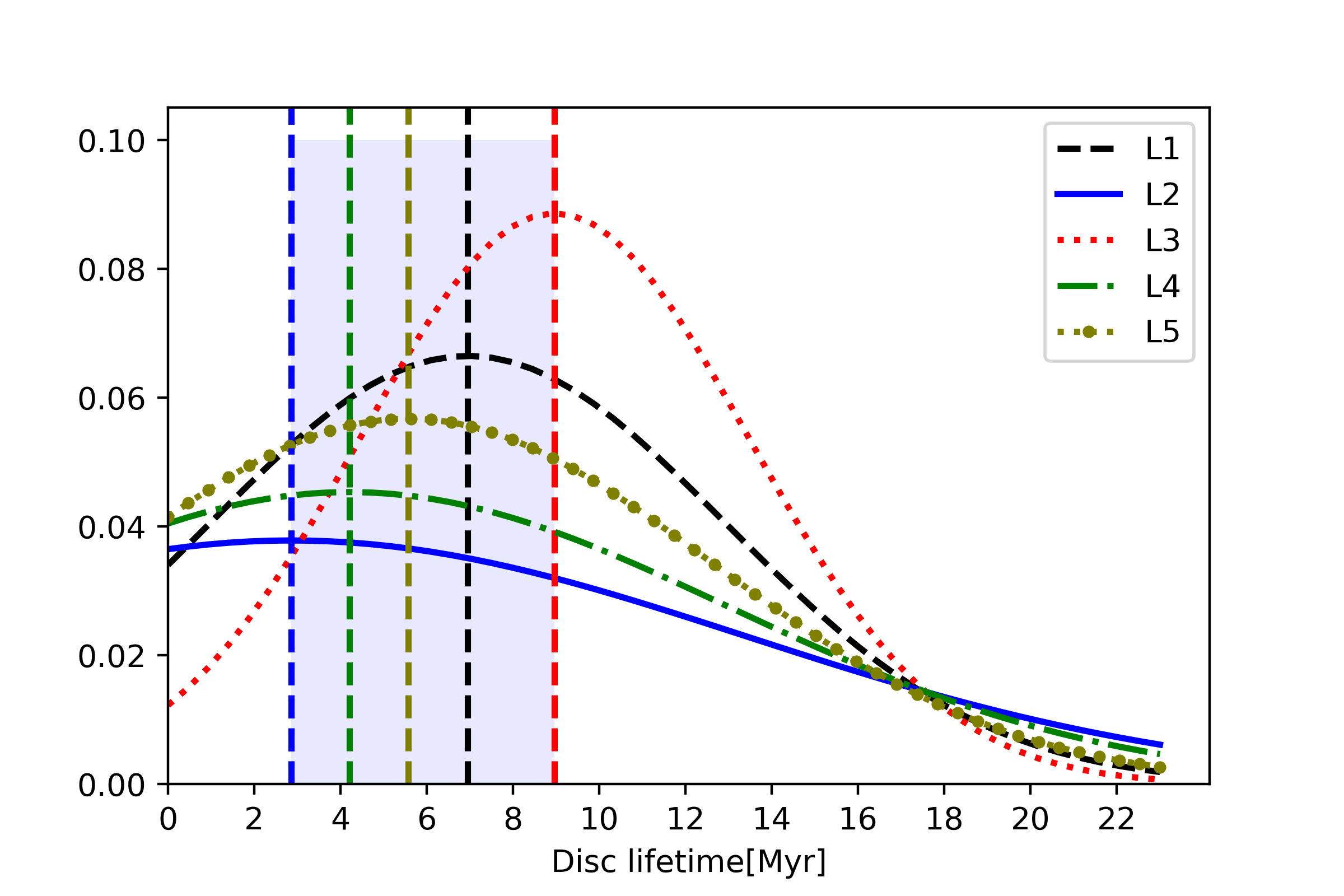}
\includegraphics[width=0.44\textwidth]{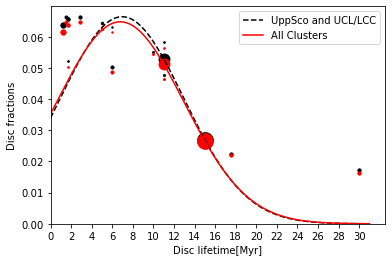}
\caption{
Gaussian-type disc lifetime distributions assuming f(t=0)= 100\%. Top: Schematics of determining the best-fit curve for model L1. Middle: Best-fits for data models L1 -- L5 for data given in Table \ref{tab:fit_gauss}, taking only the data from Upper Sco and UCL/LCC into account. Bottom: Comparison between best-fits to Upper Sco and UCL/LLC only (black, dashed line) and best-fit including clusters with $<$ 200 pc distance (red, full line) given in \mbox{Table \ref{tab:cluster_properties}}. The fit takes the relative sample size into account. The size of the symbol indicates the number of stars in the sample. In all plots, the vertical lines indicate the corresponding median disc lifetimes.}
\label{fig:distribution_disc_lifetimes_100}
\end{figure}

In the following, we systematically test the ability of different distribution functions to represent a disc lifetime distribution. When fitting the various functions to data, one must avoid biases and selection effects in the data as much as possible. Determined median disc lifetimes are very sensitive to cluster selection \citep{Michel:2021,Pfalzner:2022}. In particular, including many distant clusters in the sample leads to shorter apparent disc lifetimes. The reason is a selection bias towards higher-mass stars, which disperse their discs earlier than low-mass stars. Therefore, we consider only disc fractions of nearby ($<$ 200 pc) clusters and co-moving groups as given in \citet{Pfalzner:2022}.

We consider the different statistical significance of the various data points. Disc fractions determined for clusters containing only a few dozen stars are intrinsically more prone to errors than those from clusters consisting of thousands of stars. In most figures, the size of the symbol scales with the number of stars used to determine the disc fraction to illustrate that we account for the statistical significance of the observational data points (see, for example, Fig.\ref{fig:distribution_schematic} bottom), In the following we refer to this as sample L1++, it contains the clusters (Cham I and Cham II \citep{Galli:2021_1}, Ophiuchus and Taurus \citep{Manzo:2020}, Lupus-on-cloud and Lupus-off-cloud \citep{Spezzi:2015}, Lupus and Alessi30 \citep{Galli:2021_1, Michel:2021}, $\epsilon$Cha \citep{Michel:2021, Dickson:2021}, CrA and TW Hya and $\eta$Cha \citep{Dickson:2021}, UppSco and UCL/LCC \citep{Luhman:2022,Luhman:2020}. The clusters' disc fractions can be found in Table \ref{tab:cluster_properties} in the Appendix. Our sample's most significant disc fraction data points are those of Upper Sco and UCL/LLC. Besides, all data points are affected by errors in the disc fraction and the cluster's age. These uncertainties must be considered appropriately when determining the disc lifetime distribution.

\subsection{Gaussian distribution}
First, we fit a Gaussian distribution to the observational data points of UpperSco and UCL/LLC \citep[see also][]{Pfalzner:2022b}. We use this case to illustrate the general method. The shape of a Gaussian is fully determined if two points are given. The two values we base our estimate on are the disc fractions and ages of Upper Sco \citep{Luhman:2020} and UCL/LCC \citep{Luhman:2021} (see table \ref{tab:fit_gauss}, model L1). As mentioned, we use the values for the low-mass stars only because the disc lifetime depends on the stellar mass. The fact that discs still surround 25\% of all Upper Sco low-mass stars at 11 Myr, means that 25\% of these stars have a disc lifetime longer than \mbox{11 Myr.} Equally,
9\% of UCL/LCC low-mass stars have a disc lifetime longer than \mbox{15 Myr.} In other words, 25\% of the curve under the Gaussian is at values \mbox{$>$ 11 Myr} and 9\% of the curve under the Gaussian is at values \mbox{$>$ 15 Myr} (see \mbox{Fig. \ref{fig:distribution_disc_lifetimes_100} top).} A Gaussian with a mean of \mbox{$\mu=$ 6.95} and a standard deviation $\sigma$= 6 simultaneously fulfills these two conditions.
Thus, for these conditions, we obtain a mean disc lifetime of approximately 7 Myr with an extensive spread around this mean value.

The disc fractions and the cluster ages are subject to uncertainties. We performed additional Gaussian fits exploring the combinations of extremes in disc fraction and age for Upper Sco and  UCL/LLC  (see model L2 -- L5 in table \ref{tab:fit_gauss}) to test the sensitivity of the result. Depending on the assumed cluster age and disc fraction, the mean disc lifetime varies between 5 and 8 Myr, with all distributions having a standard deviation $\sigma >$ 4.5 Myr (see Fig. \ref{fig:distribution_disc_lifetimes_100}, middle).

So far, our fit solely relies on the disc fractions of Upper Sco and UCL/LCC. In \citep{Pfalzner:2022}, we demonstrated that more distant clusters are often biased towards higher-mass stars. Therefore, we expand our larger sample to only clusters with distances $<$ 200 pc (see Table \ref{tab:cluster_properties}). In the following, this larger sample is referred to as model L1++. In Fig. \ref{fig:distribution_disc_lifetimes_100} bottom, we show a comparison between the best fit to this larger cluster sample and the distribution based on  Upper Sco and UCL/LCC alone. The fit was obtained, taking the sample size into account.  It can be seen that both distributions resemble each other closely. Thus, given the still large uncertainties in the data, both distributions are equally suitable.

We test the fit quality $\Delta D$ by calculating the mean least square distance of the data points to the best-fitting curve. When summing these errors, we weigh the data points according to the number of stars $n_i$ used to determine the disc fraction,
\be
  D = \sum_i  \frac{n_i}{\sum_i n_i} \sqrt{(\hat{t}_i -  t_j)^2 + (\hat{f}_i - f_j)^2}, 
  \nonumber
\ee
where $\hat{t}_i$ are the cluster ages, $\hat{f}_i$ the cluster disc fractions; $t_j$ and $f_j$ represent the points on top of the distribution plotted from initial parameter L1. Fig. \ref{fig:distribution_disc_lifetimes_100} bottom shows the best-fit distributions, considering the additional data points. We see that the change is only marginal. The reason is the relatively small number of stars in these clusters, giving them a low weight in the fit. 
The best Gaussian disc lifetime distribution is
\be 
T(t)  =  \frac{1}{\sqrt{12 \pi} } exp  \left[-1/2 \left(\frac{t - 6.95}{6}\right)^2\right].
\ee
However, there exists a potential problem with a Gaussian distribution. There exists a  cut-off of the distribution at \mbox{$t$= 0 Myr,} which can be interpreted as a considerable fraction of stars having negative disc lifetimes. Two conclusions exist: (i) the disc fraction distribution has a non-Gaussian shape, and (ii) the initial disc fraction is $<$ 100\%. In the following, we investigate both options and their combination.

\begin{table*}
    \setlength{\tabcolsep}{2pt}
        \caption{Parameters for the different distribution types}
        \centering\begin{tabular}{lcccccccccccc}
        
        \hline
& \multicolumn{4}{c}{$f_{init}$ = 1.0} &   \multicolumn{4}{c}{$f_{init}$ = 0.8} & \multicolumn{4}{c}{$f_{init}$ = 0.65} \\ 
        \hline
Type & Parameters & $t_d$ & $\sigma(t_d)$  &  $t_{max}$ & Parameters & $t_d$ &  $\sigma(t_d)$   &  $t_{max}$ &  Parameters & $t_d$ & $\sigma(t_d)$  &  $t_{max}$\\
        \hline

Gaussian      & \multicolumn{1}{p{2.5cm}}{\centering{\textbf{$\sigma = 6.0$} \\ \textbf{$\mu   =  6.95$}}} & 6.95 & 6  &  6.95 & \multicolumn{1}{p{2.5cm}}{\centering{\textbf{$\sigma = 5.07$} \\ \textbf{$\mu   =  9.55$}}}  & 9.55 & 5.07  &  9.55 & \multicolumn{1}{p{2.5cm}}{\centering{\textbf{$\sigma = 5.04$} \\ \textbf{$\mu   =  9.52$}}} & 9.52 & 5.04  &  9.52 \\
        
Gamma      & \multicolumn{1}{p{2.5cm}}{\centering{\textbf{$\alpha = 3.53$} \\ \textbf{$\beta = 0.41$}}} & 8.5 & 20.6  &  5.99 & \multicolumn{1}{p{2.5cm}}{\centering{\textbf{$\alpha = 4.47$} \\ \textbf{$\beta = 0.47$}}}  & 9.4 & 19.9  &  7.31 & \multicolumn{1}{p{2.5cm}}{\centering{\textbf{$\alpha = 5.69$}  \\ \textbf{$\beta = 0.55$}}} &  10.32 & 18.7 & 8.44\\

Log-normal & \multicolumn{1}{p{2.5cm}}{\centering{\textbf{$\sigma = 0.46$} \\ \textbf{$\mu   =  2.08$}}}  &  8.9   &  19.4     & 6.43 & \multicolumn{1}{p{2.5cm}}{\centering{\textbf{$\sigma = 0.43$} \\ \textbf{$\mu   =  2.19$}}}     &  9.8   &  19.2    & 7.40 & \multicolumn{1}{p{2.5cm}}{\centering{\textbf{$\sigma = 0.39$} \\ \textbf{$\mu   =  2.28$}}} & 10.59 & 18.48 & 8.42 \\

Weibull    & \multicolumn{1}{p{2.5cm}}{\centering{\textbf{$k = 1.78$} \\ \textbf{$\lambda   = 9.15$}}}  &  8.1   &  22.4      & 5.80  & \multicolumn{1}{p{2.5cm}}{\centering{\textbf{$k = 2.03$} \\ \textbf{$\lambda  = 10.21$}}}    & 9.0    &  21.7   & 7.20 & \multicolumn{1}{p{2.5cm}}{\centering{\textbf{$k = 2.34$} \\ \textbf{$\lambda   = 11.22$}}} &  9.93 & 20.29 & 8.83 \\

Log-logistic & \multicolumn{1}{p{2.5cm}}{\centering{\textbf{$\alpha = 8.31$} \\ \textbf{$\beta   = 3.92$}}} &  9.3   &  24.8     & 7.21 & \multicolumn{1}{p{2.5cm}}{\centering{\textbf{$\alpha = 9.08$} \\ \textbf{$\beta   = 4.12$}}}    &  10.0   &  25.5  & 8.04  & \multicolumn{1}{p{2.5cm}}{\centering{\textbf{$\alpha = 9.88$} \\ \textbf{$\beta   = 4.38$}}} & 10.78 & 25.13 & 8.89 \\

Beta & \multicolumn{1}{p{2.5cm}}{\centering{\textbf{$p = 3.02$} \\ \textbf{$q   =  32.92$}}}  &  8.4   &  20.8       &  5.97  & \multicolumn{1}{p{2.5cm}}{\centering{\textbf{$p = 3.85$} \\ \textbf{$q   =  37.32$}}}     &  9.3   & 20.1    & 7.27 & \multicolumn{1}{p{2.5cm}}{\centering{\textbf{$p = 4.92$} \\ \textbf{$q   =  43.12$}}} & 10.25 & 18.76 & 8.52 \\
 \hline
 
              \end{tabular}
        \label{tab:fit_distributions}
        \begin{flushleft}
        \tablecomments{Column~1 indicates the distribution type, $f_{init}$  stands for the assumed initial disc fraction,  $t_d$, the median disc lifetime, $\sigma(t_d)$ the standard deviation, and $t_{max}$ the maximum of the distribution. The corresponding $\Delta D $-values, describing the quality of the fit, can be found in table \ref{tab:D-values} in the Appendix.}
        \end{flushleft}
\end{table*}

\subsection{Tested types of distribution}

Positively skewed distributions might better represent the disc lifetime distribution while avoiding the problem of negative disc lifetimes. Therefore, we tested the 
\be 
Gamma:T(x, \alpha, \beta)  & = & \frac{\beta^{\alpha}}{\Gamma(\alpha)} x^{\alpha - 1}  e ^{-\beta x}, \nonumber \\
Log-normal: 
T(x, \sigma, \mu) & = & \frac{1}{\sqrt{2 \pi} \sigma x} exp \left( - \frac{[ln(x) - \mu]^2}{2 \sigma^2} \right),  \nonumber \\
Weibull: T(x, k, \lambda) & = &  \frac{k}{\lambda} \left(\frac{x}{\lambda} \right)^{k-1} exp \left(- \left(\frac{x}{\lambda} \right)^k \right),
\nonumber \\
Log-logistic: T(x, \alpha, \beta) &=&  \frac{(\beta/\alpha)(x/\alpha)^{\beta-1}}{(1+ (x/\alpha)^\beta))^2}, \nonumber \\
Beta:  
T(x, p, q) & = &\frac{\Gamma(p+q)}{\Gamma(p)\Gamma(q)}x^{p-1}(1-x)^{q-1} \nonumber
\ee
distributions as possible candidates for the disc lifetime distribution. See Fig. \ref{fig:distribution_schematic} middle for a schematic representation of these distributions.

\begin{figure*}[t]
\centering\includegraphics[width=0.47\textwidth]{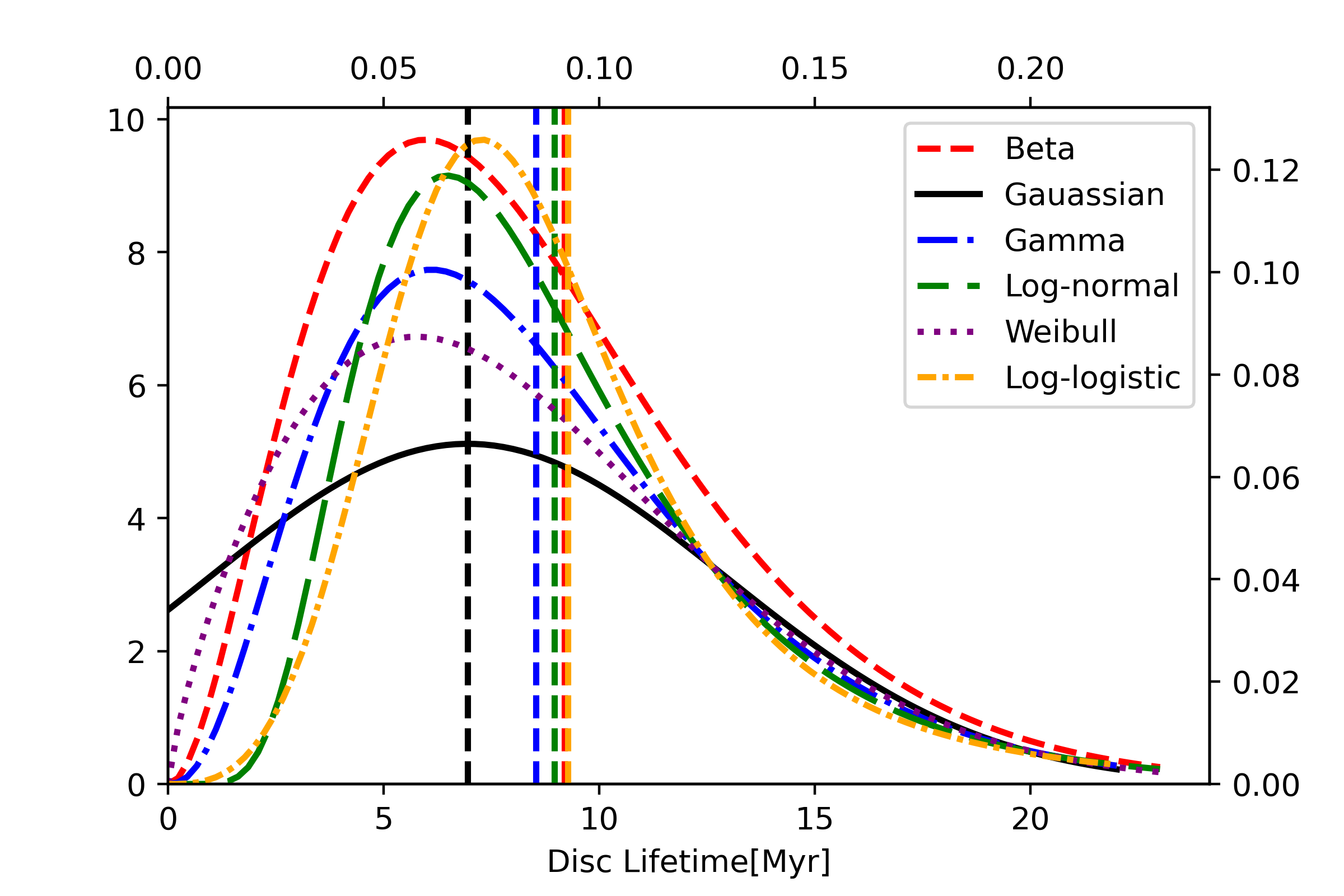}
\includegraphics[width=0.47\textwidth]{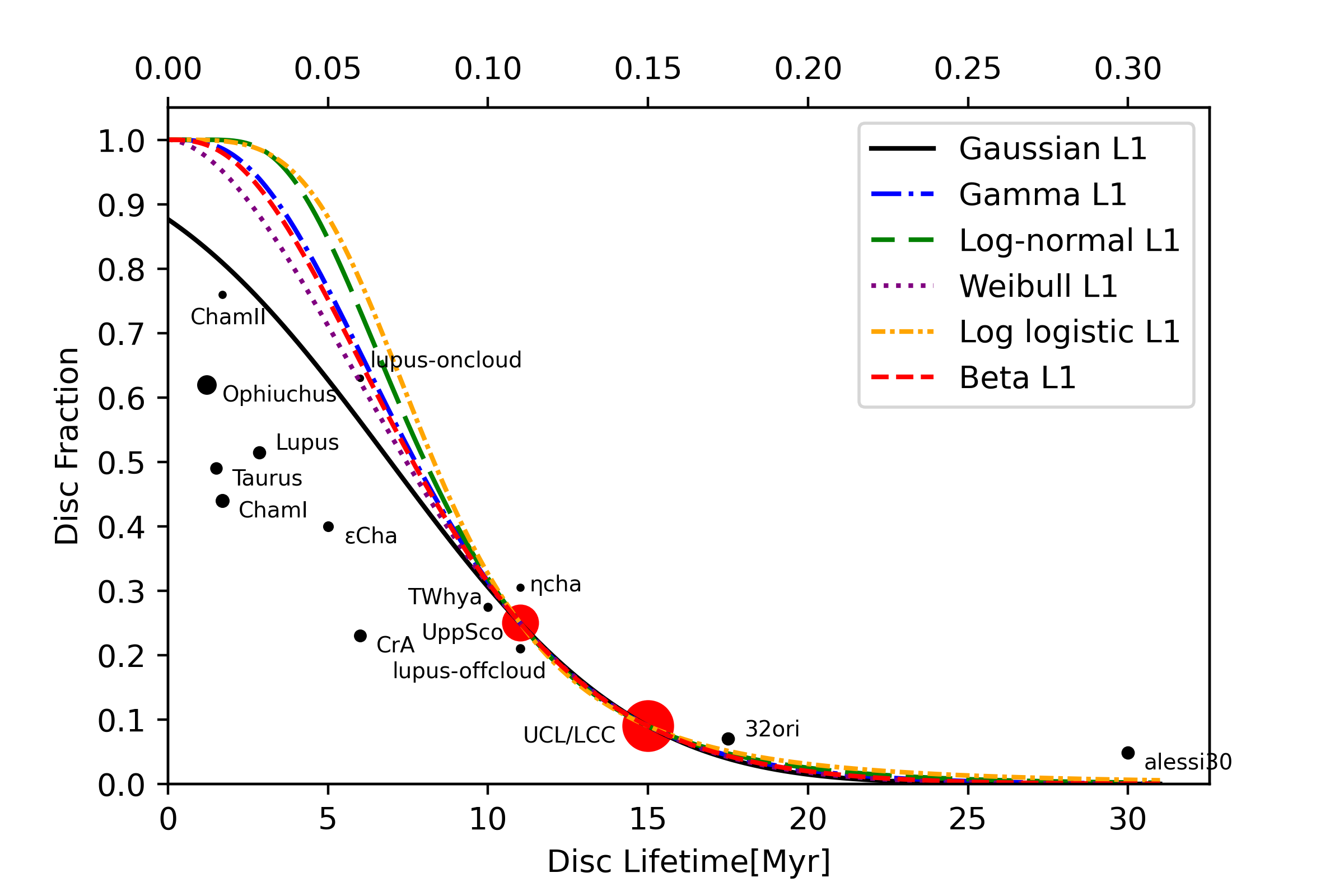}
\includegraphics[width=0.47\textwidth]{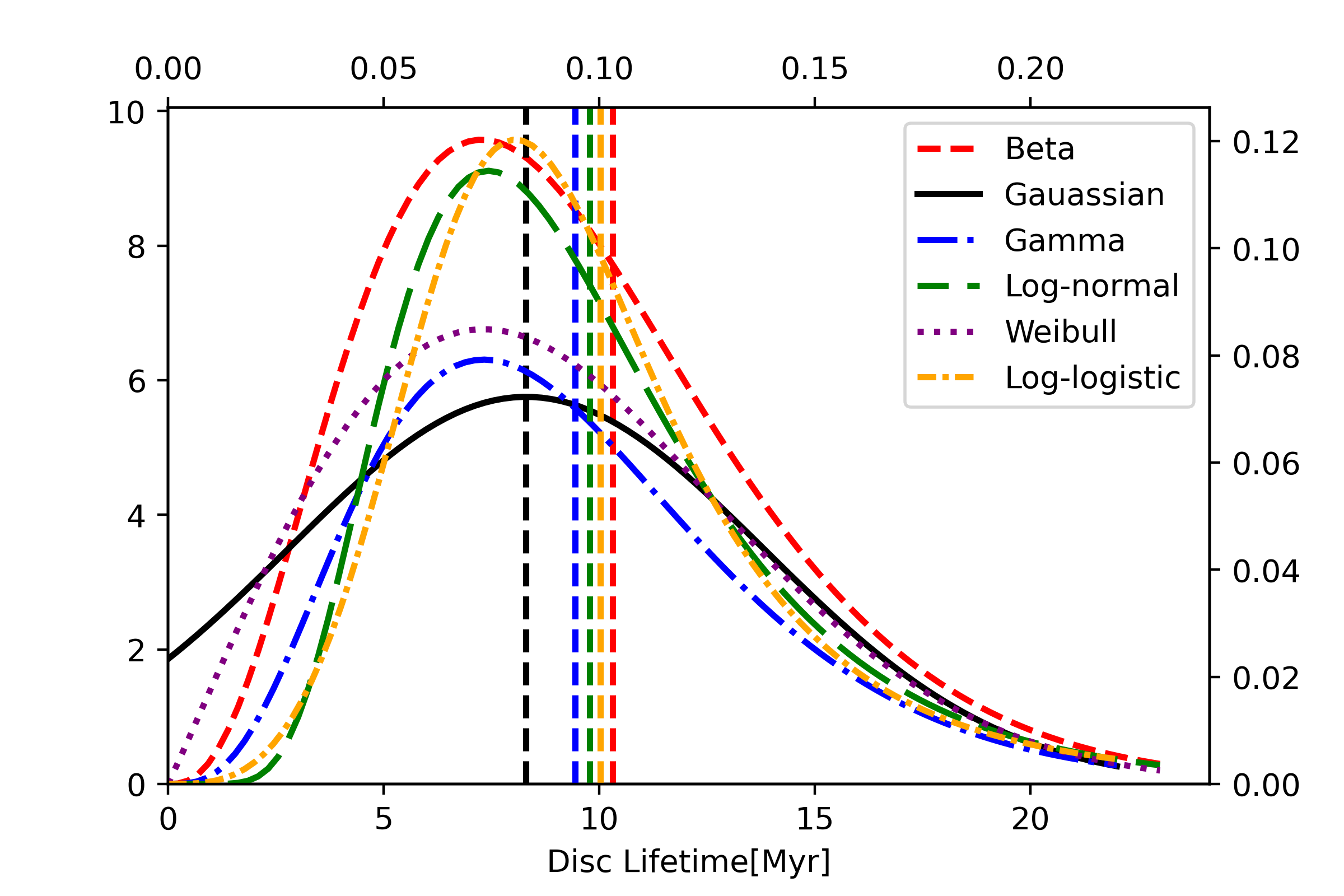}
\includegraphics[width=0.47\textwidth]{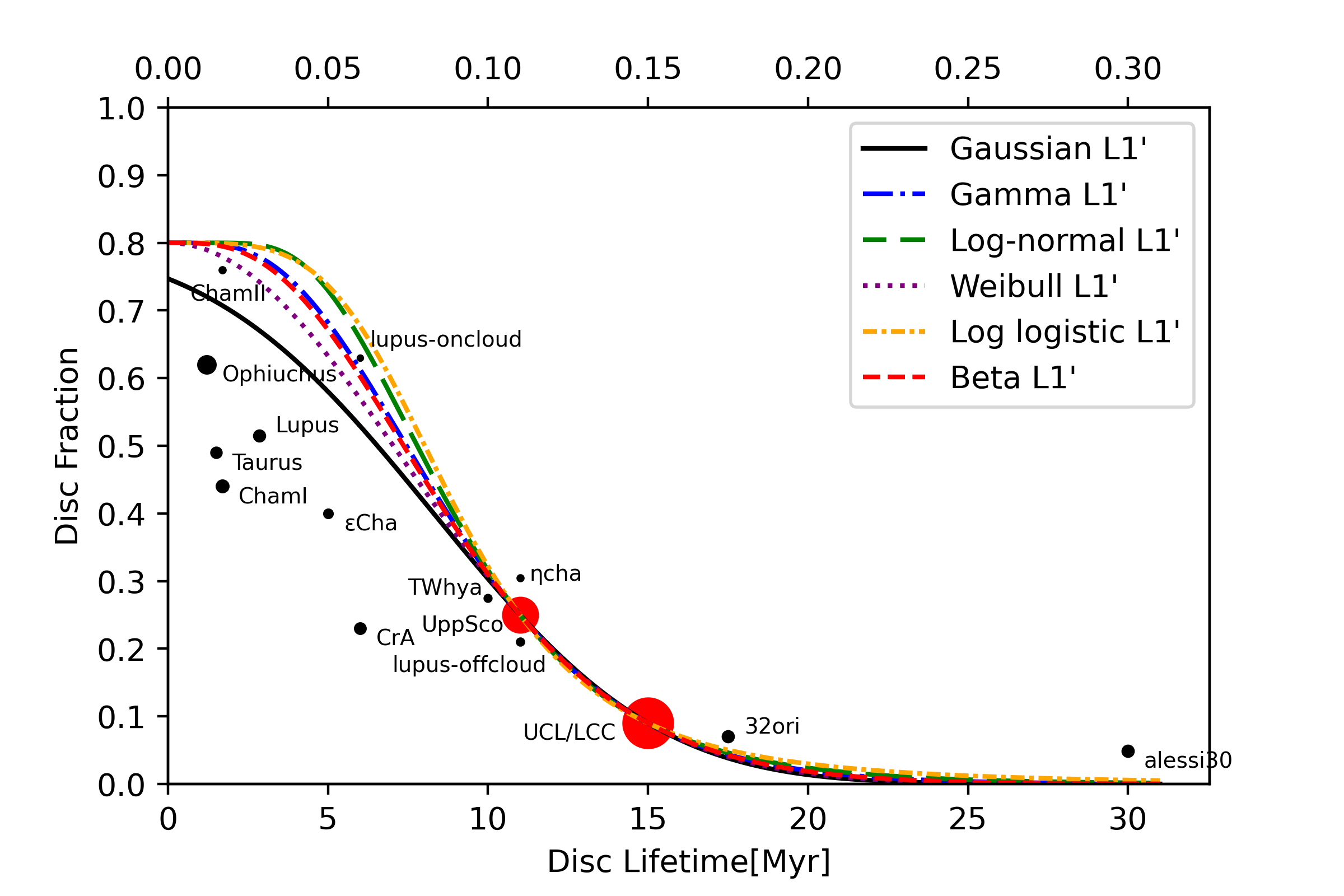}
\includegraphics[width=0.47\textwidth]{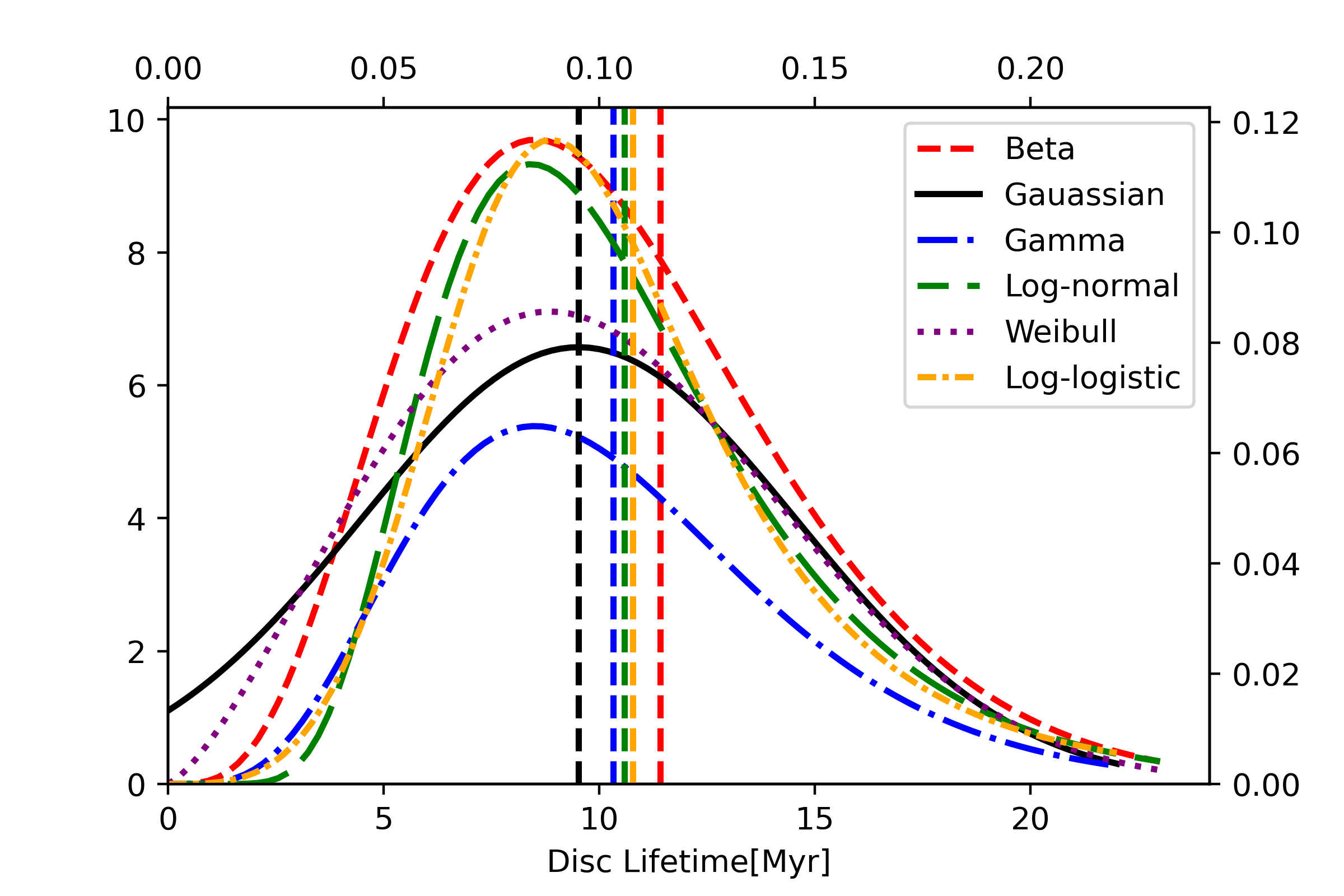}
\includegraphics[width=0.47\textwidth]{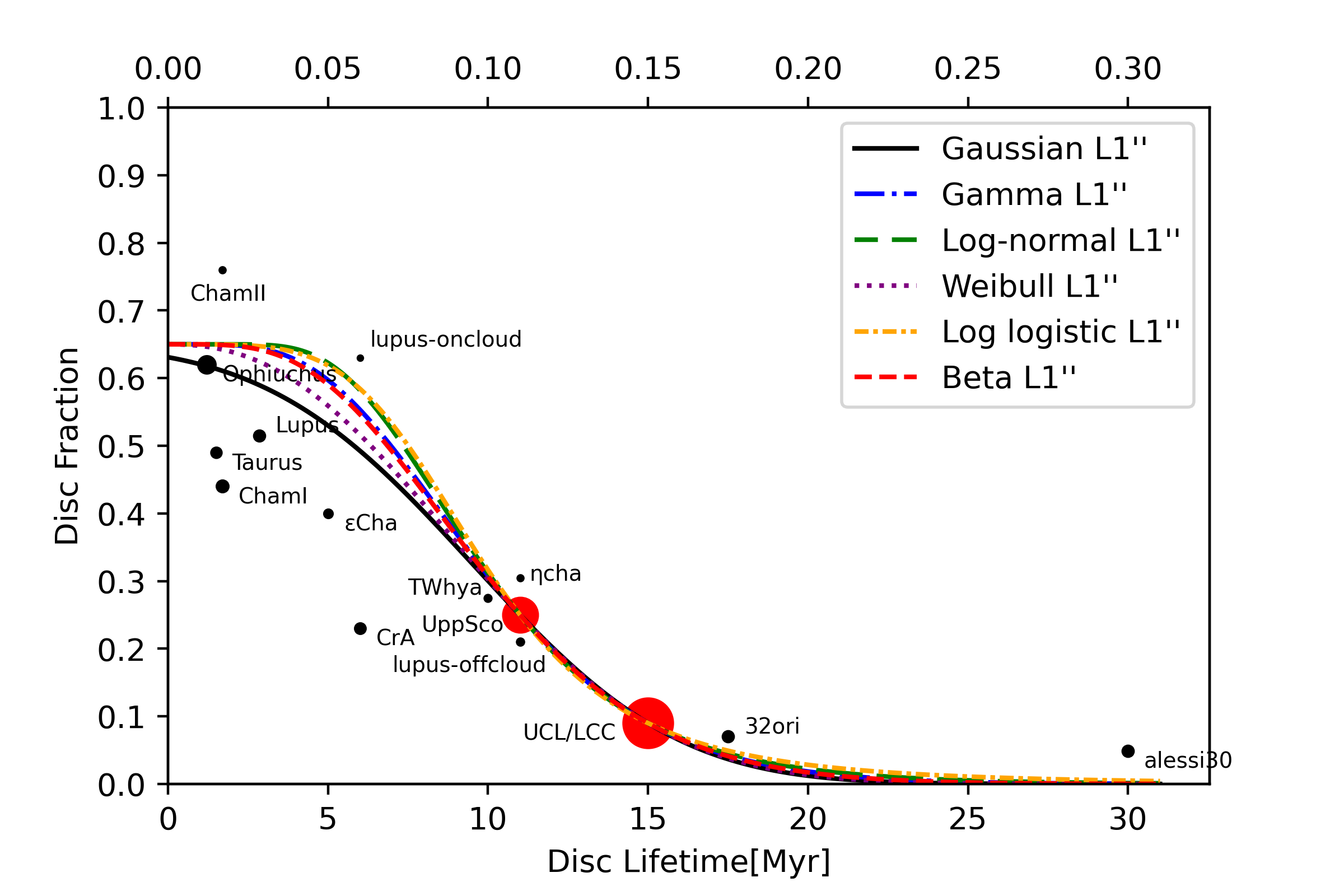}
\caption{Left column: Disc lifetime distributions assuming a Gamma, Log-normal, Weibull, Log-logistic, and Beta distribution shape. These fits are based on the L1 data for Upper Sco and UCL/LLC. Right column: Corresponding disc fraction vs. cluster age plots assuming the various distributions. The top row shows the results assuming a 100\% initial disc fraction, whereas the middle row illustrates the case of an initial disc fraction of 80\% and the bottom row for an initial disc fraction of 65\%. The vertical lines indicate the corresponding median disc lifetime. }
\label{fig:distribution_comparison_100}
\end{figure*}

We determine the best for all these distribution types to the data of set L1 for Upper Sco and UCL/LLC (see Fig. \ref{fig:distribution_comparison_100}, left panel). Table \ref{tab:fit_distributions} provides the corresponding parameters of these best fits. The equivalent values for models \mbox{L2 -- L1++} can be found in Table \ref{tab:fit_parameters_Gauss} in the Appendix. In a second step, we determined the deviation from the additional disc fraction data points $\Delta D $. These values are given in table \ref{tab:D-values} in the appendix.

All these distributions start at $f(t$=0) = 0, show a distinct maximum and have a positively skewed shape. Thus, they seem a more realistic approach than an exponential function. 
All distributions give remarkably similar results for the disc lifetime distribution. Thus, using any of these could be justified given the still-existing uncertainties in disc fractions and ages. 
The mean disc lifetime of these distributions lies in the range 8.1 -- 9.3 Myr.     Thus, these values agree with the disc lifetime of $\approx$ 8 Myr found by \cite{Michel:2021}. For left-sewed distributions, the median disc lifetime and the maximum in the distribution differ. This maximum corresponds to the time at which most stars shed their discs.
This maximum lies in the 5.8 Myr -- 7.21 Myr range for the best-fit distributions. 

Of all investigated distribution types, we favour the following Weibull distribution,
\be 
T(t)  =  91.57 t^{0.78} exp (-51.44 t^{1.78}),
\ee

for two reasons: First, it gives the smallest $\Delta D $ value. Second, converting these disc lifetime distributions back into a disc fraction vs. cluster age plot (see Fig. \ref{fig:distribution_comparison_100}, right panel) shows the lowest discrepancy to the disc fractions younger than  5 Myr. However, at these young ages, the observed disc fractions are lower than expected from any of these distributions. One reason could be the assumption that all stars are initially surrounded by a disc. However, this assumption may not be justified.

\subsection{Role of initial disc fraction}

\citet[][]{Michel:2021} obtain much better fits to their data when assuming initial disc fractions of 80\% and 65\% rather than 100 \%. They attribute the reason for a \mbox{$<$ 100\%} disc fraction to the presence of binary stars. Other reasons for the initial disc fraction being less than 100\% could be that some stars are born without a disc or lose their disc rapidly \mbox{($<$ 0.5 Myr)} \citep[][]{Richert:2018}. 

Another reason could be the age spread in the cluster. For instance, star formation in the star-forming region of Taurus has been ongoing for at least 1 -- 3 Myr  \citep{Kenyon:1995}. Similarly, across Orion, for discs in isolation, the spread in disc mass is likely related to an age spread. \citep{Terwisga:2022}. Thus, even if all stars were initially surrounded by a disc, at $t$ = 0 Myr, the oldest stars of the distribution may have already lost their discs, leading to 
$f_d$(t=0) $<$ 100\%. 

Assuming a lower initial disc fraction affects the shape of the disc lifetime distribution. The two bottom rows of \mbox{Fig. \ref{fig:distribution_comparison_100}} show the best-fit solutions assuming an initial disc fraction of 80\% and 65\%, respectively. Like before, all distributions start at $f(t$=0) = 0, show a distinct maximum and have a positively skewed shape. Assuming a lower initial disc fraction leads to the mean disc lifetime of these distributions being slightly higher. For $f_{init}$ = 80\%, the mean disc lifetimes are 9.0 -- 10 Myr. The $D$ is lower D due to the match to the disc fractions of clusters with ages $<$ 5 Myr being better. Like before, a Weibull distribution gives the best match. The best-fit parameters are again listed in table \ref{tab:fit_distributions}.

\begin{figure}
\includegraphics[width=0.48\textwidth]{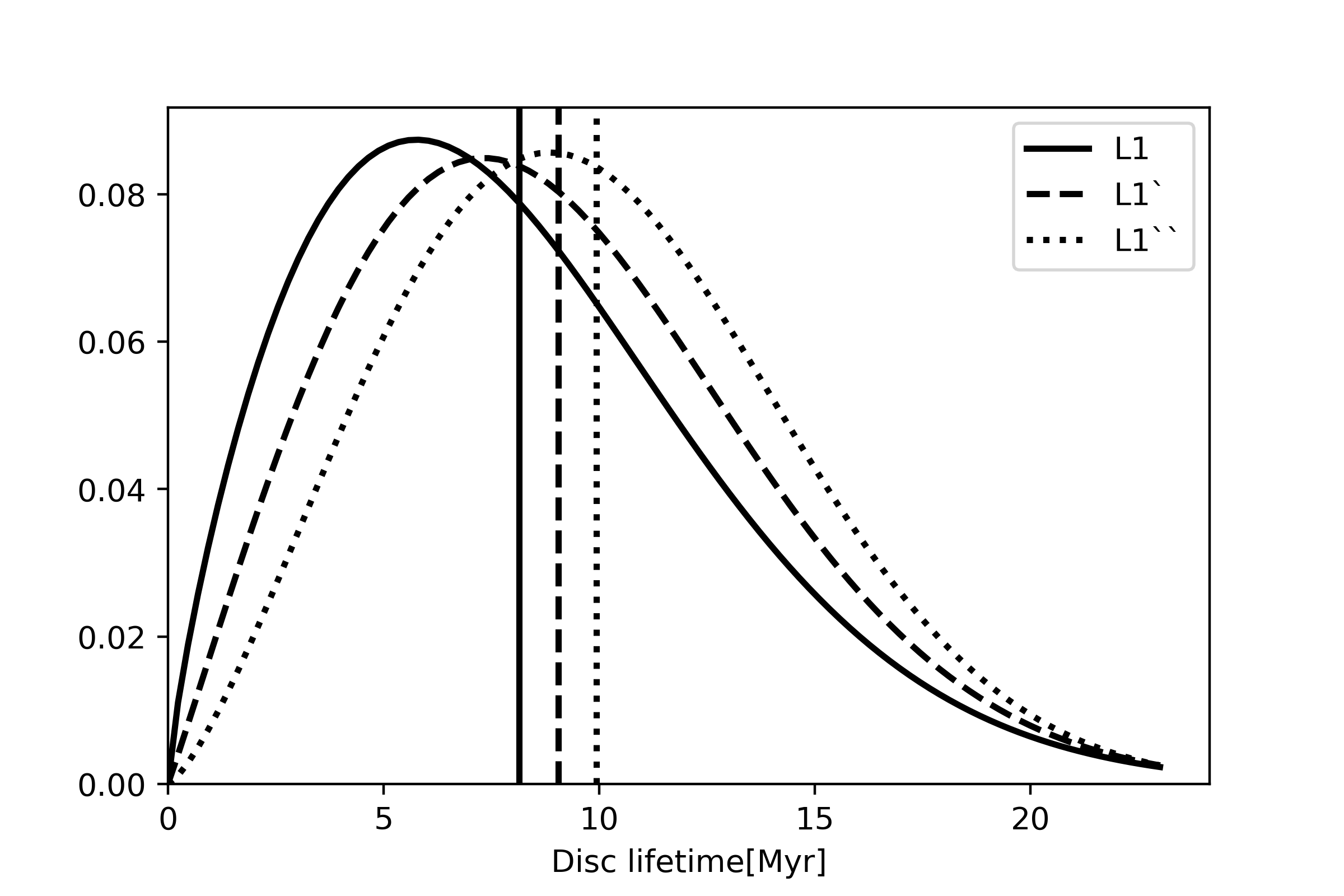}
\includegraphics[width=0.48\textwidth]{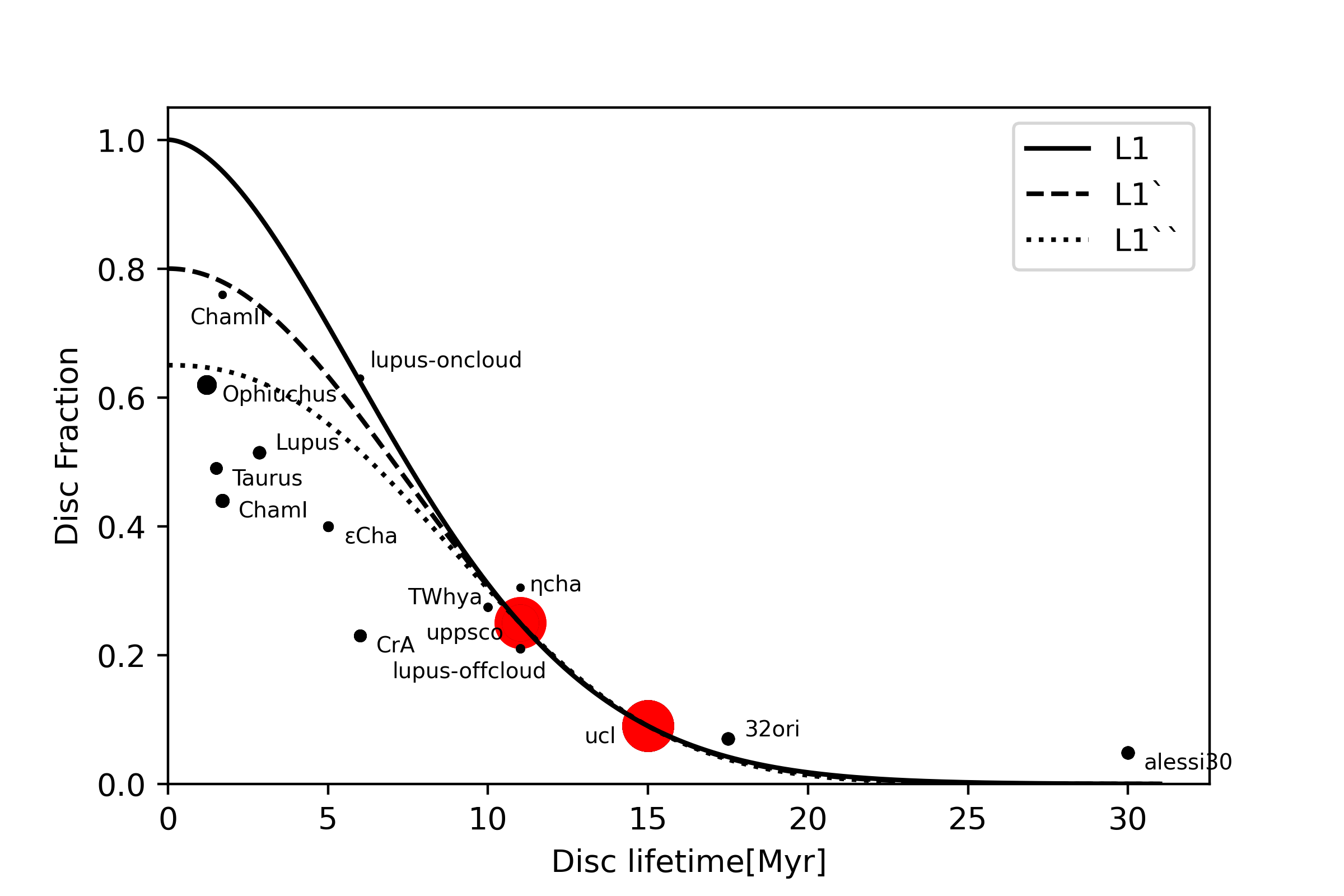}
\caption{Comparison of Weibull distributions assuming a  $f_{init}$ = 1.00 (full line),  $f_{init}$ = 0.80 (dashed line) and  $f_{init}$ = 0.65 (dotted line) initial disc fraction.}
\label{fig:distribution_Weibull}
\end{figure}
%

%

\begin{figure}
\centering
\includegraphics[width=0.48\textwidth]{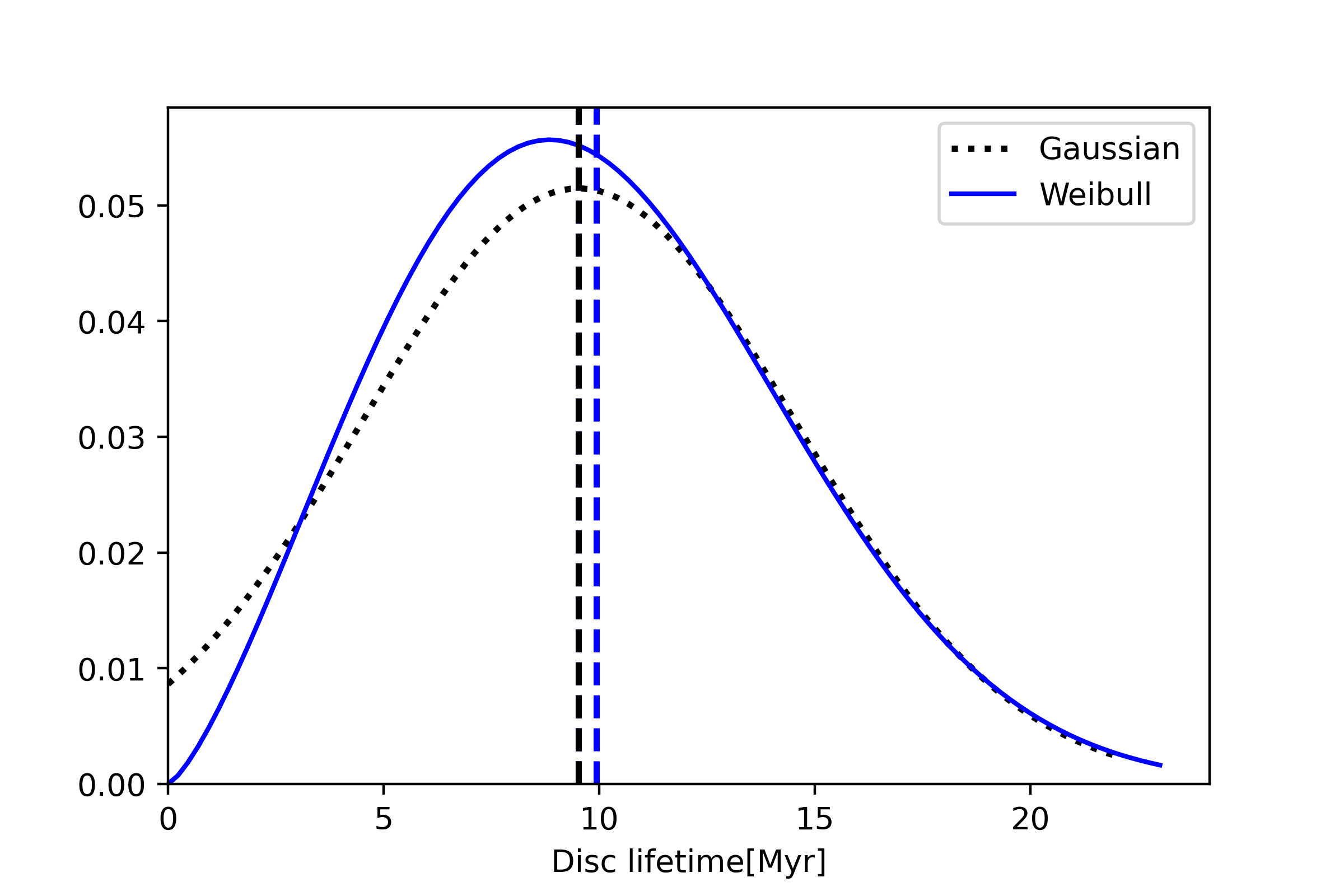}
\includegraphics[width=0.48\textwidth]{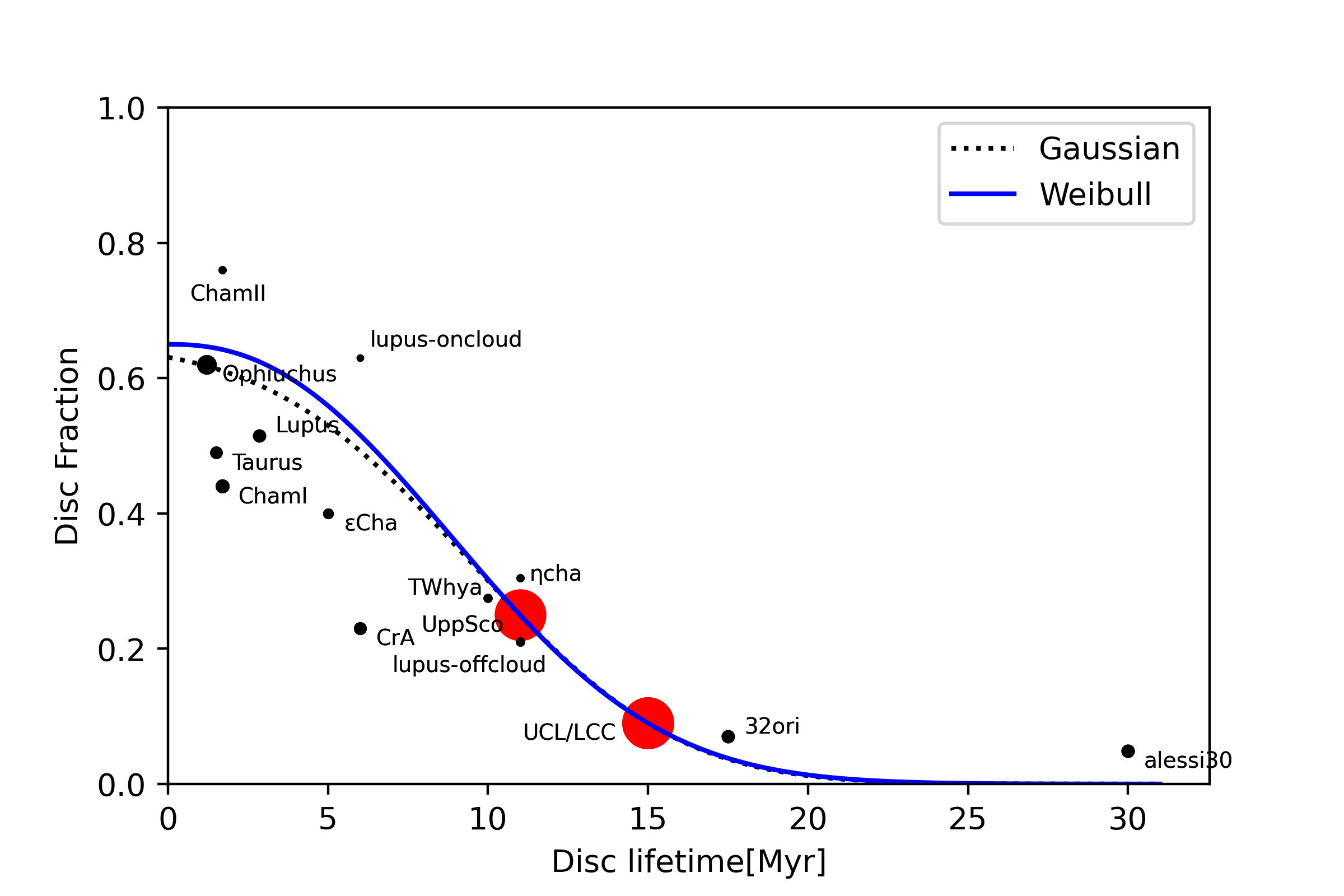}
\caption{Comparison between Gaussian (dotted line) and Weibull distribution. Top: Distribution, Bottom: Corresponding disc fraction vs. cluster age diagram.}
\label{fig:Gauss_Weibull}
\end{figure}

Figure \ref{fig:distribution_Weibull} shows a comparison of the Weibull-type distributions assuming an initial disc fraction of 100\% (full line), 80\% (dashed line) and 65\% (dotted line) and also the corresponding curves in the disc faction vs. cluster age illustration. It can be seen that lower initial disc fractions result in the maximum disc destruction shifting to later times. From $t_max$ (100\%) = 5.8 Myr in the case of 100\% initial disc fraction, it increases to $t_{max}$ (80\%)= 7.2 Myr and $t_{max}$ (65\%) = 8.83 Myr for lower initial disc fractions. Correspondingly, the median disc fraction increases from $t_d$(100\%) = \mbox{8.14 Myr} to $t_d$(65\%) = 9.93 Myr. Nevertheless, the curve for 65\% would be better to reconcile with the observed disc fractions for clusters $<$ 4 Myr. The $D$-values decrease from $D(10\%)$ = 0.077 to $D(65\%)$ = 0.034.  The best fit for Weibull assuming an initial disc fraction of 65\% is
\be 
T(t)  =  0.209 \left(\frac{t}{11.22} \right)^{1.34} exp \left(- \frac{t}{11.22}^{2.34} \right),
\ee
Surprisingly, this distribution differs only slightly from the Gaussian distribution for 65\% initial disc fraction (see Fig,\ref{fig:Gauss_Weibull}), which is given by 

\be 
T(t)  =  \frac{1}{\sqrt{12 \pi} } exp  \left[-1/2 \left(\frac{t - 6.95}{6}\right)^2\right].
\ee
Thus, for all practical purposes, both heuristic fits could be used.

\section{Current Limitations}

The derived disc lifetime distribution should be regarded as a first iteration step. The challenge is the scarcity of nearby clusters in the age range 5 Myr -- 15 Myr. Our solution was to use the disc fraction of Upper Sco and UCL/LLC  as anchor points for determining the disc lifetime distribution. However, 
there has been a long discussion in the literature concerning the age of Upper Sco. The age determinations in the literature from the last decades fall broadly around
two estimates: 5 Myr (\citep{de_Geus:1989A&A...216...44D,Preibisch:2002} and 10–12 Myr \citep{Pecault:2012,Sullivan:2021}. Here, we used the latter age.
\\
The question is whether it is realistic to attribute Upper Sco and UCL/LLC specific ages. Recent studies of Upper Sco by \citet {Ratzenboeck:2023} based on new GAIA data show that Upper Sco's stellar population consists of up to nine subclusters aged between 3 Myr and 19 Myr. However, determining individual age is also challenging and can be affected by stellar rotation, stellar spots, and photometric variability. The differences in ages of the individual sub-clusters can explain the wide age spread and the conflicting results in earlier studies. Similarly, the UCL/LCC area can be divided into separate subclusters of different ages. They find that the entire Sco-Cen area is dominated by a brief period of intense star and cluster formation rate about 15 Myr ago, slowly declining since this burst. 
\\
As the sub-clusters in Upper Sco and ULC/LLC have ages roughly in the range of 3 Myr - 20 Myr, one can obtain disc fractions in this age range. Once these disc fractions are published, one can obtain the next better iteration of the disc lifetime distribution with the method described here.

Determining the initial disc fraction at $t$= 0 is a challenging task. To better understand the influence of the age spread in clusters on disc fractions, it is crucial to make significant efforts in this area. One way of achieving this is by determining the age of individual stars in clusters. This will enable us to put better observational constraints on the $t$= 0 disc fraction in the future.

\section{Reasons for the large width of the distribution}
The obtained disc lifetime distributions are all very broad ($\sigma >$ 6 Myr). It is intriguing to consider the reasons for the vast differences in disc lifetime, ranging from just 1 Myr to over 20 Myr. Several factors have been identified as potential influencers of individual disc lifetime, including stellar mass, initial disc mass, environment, planet formation, and stellar binarity. 

\subsection{Stellar mass}
Stellar mass definitely influences the disc lifetime considerably \citep[e.g.,][]{Carpenter:2006,Roccatagliata:2011,Yasui:2014,Ribas:2015,Richert:2018,Marel:2021,Pfalzner:2022}.  However, we separated this element by restricting our study to low-mass stars of spectral type M3.7 - M6 following the classification of \citet{Luhman:2022}. They estimate that the contamination is $\approx$ 1$\%$ and the completeness is  $\approx$ 90\% at spectral types of $\leq$M6–M7 for the populations with low extinction in Upper Sco UCL/LCC. Based on a comparison to deeper spectroscopic surveys of Upper Sco, they find that the completeness of the Gaia sample of candidates in that region begins to decrease at spectral types later than M7 ($\approx$ 0.06 \MSun). In this study, all of the populations in Sco-Cen peak for spectral types near M4–-M5. This indicates that the populations of UpperSco and UCL/LLC share similar characteristic masses for their initial mass functions. 

The additional clusters used for model L1++ are not restricted to this mass range, because they contain much fewer(apart from Ophiuchus, $<$ 200) stars than UpperSco and UCL/LLC. Due to the IMF, these clusters are dominated by M- and L-type stars, but some contain also a few early-type stars. Thus, the presence of higher stars ($\geq$ M3) might lower the disc fractions, as the disc lifetime tends to decrease for higher stellar masses. However, due to the dominance of M- and K-type stars, these few early-type stars do not the disc fractions in a severe way \citep{Marel:2023}. Nevertheless, it might influence the estimates of the initial disc fraction and should be investigated in this context further. 

Above, we showed that the spread in disc lifetimes remains even when concentrating on this specific stellar mass range. We investigate the role of stellar mass on the disc lifetime distribution in detail in Pfalzner \& Dincer (in preparation).

\subsection{Initial disc properties}

The masses of discs surrounding similar-aged stars can vary by orders of magnitude \citep[e.g. ][]{Pasccuzi:2016,Ansdell:2017,Eisner:2018,Cieza:2019,Cazzoletti:2019,Tobin:2020,Tychoniec:2020,Grant:2021,Appelgren:2023}, even if the stars are of the same mass. This diversity in disc masses was found in the gas and dust masses alike. Similarly, it has been shown that the total disc masses in a cluster decrease with cluster age \citep{Andrews:2020}. Thus, the broad disc lifetime distribution could reflect the differences in initial disc masses. Discs of low mass would be faster accreted and more easily dispersed than high-mass discs. This answer is only partially satisfying because it replaces the question of the origin of the disc lifetimes with that of the diversity of disc masses. Besides, some old discs ($>$ 10 Myr) still have relatively high disc masses.

\citet{Michel:2021} suggest that the degree of structuring is crucial for the disc lifetime. They propose that two classes of discs -- structured and unstructured -- exist. The unstructured discs lose their dust masses quickly, whereas, in structured discs, dust traps allow for forming planetesimal belts at large radii, slowing down mass loss. The observation that structured protostellar discs are brighter and more extensive than unstructured discs supports this suggestion {\bf \citep{Michel:2021}.} However,  the problem of resolving structure in low-mass discs might introduce a bias here.

\subsection{Environment}

The environment can also influence the disc lifetime \citep[for example,][]{Adams:2006}. In regions of high stellar density, close flyby can truncate discs and, in extreme cases, even annihilate them  \citep[for example,][]{Olczak:2010,Vincke:2016,Winter:2020,Cuello:2023}. If the environment is dense, relatively gas-free and contains high-mass stars, the radiation of these high-mass stars can destroy the discs around stars in their vicinity \citep[for example,][]{Owen:2010, Sellek:2020,Concha:2021,Parker:2021,Winter:2022}. However, the main clusters in our study, Upper Sco and UCL/LLC, have a relatively low stellar density. Currently, stellar flybys and external photo-evaporation likely play only a minor role in disc lifetime diversity. However, both clusters might have once been much denser
\citep{Pfalzner:2013}. Thus, it can not be excluded that the once denser environment might be responsible for the short end of the disc lifetime distribution.

Even in low-density environments, about 10\% of stars are likely to experience close flybys
\citep{Pfalzner:2022}. Thus, the environment may lead to premature disc dispersal in about 10\% of stars.

\subsection{Planet formation}

If the planet formation is very efficient in converting the disc's gas and dust into one or more planets, this might reduce the disc mass so much that the remnant material is quickly dispersed. Such a process of premature disc dispersal would be particularly efficient if a massive gas giant is formed. Planet formation by gravitational instability would be one way to form planets fast \citep{Boss:1997}. Another process that could trigger a planet's early and fast formation in a disc is the presence of interstellar objects as seeds for planet formation \citep{Pfalzner:2021}.

\subsection{Stellar binarity}

The presence of a companion star influences the disc properties and frequency. A companion can prevent disc formation, limit its lifetime, or truncate it \citep{Harris:2012}. Besides, binaries can clear a cavity of 2--5 times the binary separation \citep{Arty:1994,Hirsh:2020,Marel:2021}. These effects lead to lower disc fraction around close binary stars compared to single stars \citep{Kraus:2012}. Besides, discs surrounding binary stars tend to be smaller \citep{Akeson:2019}. The binarity of stars is known to increase with stellar mass
\citep{Raghavan:2010} and still changes during the first 10 Myr within the host cluster of the stars 
\citep{Korntreff:2012, Kaczmarek:2011}. Thus, significant uncertainties exist concerning the binary fraction around low-mass stars younger than 10 Myr.

This study was limited to low-mass stars, which typically have a binary fraction of $\approx$ 20 \%. If close binaries prohibit disc formation in some cases, then stellar binarity may play a role in the initial disc fraction. If binarity is the primary mechanism for reducing the initial disc fraction, then the initial disc fraction should be smaller for higher-mass stars. Determining whether the initial disc fraction declines with stellar mass should be able to quantify the relevance of binarity in this context. 

Binarity may also contribute to the portion of discs with relatively short disc dissipation times. We can estimate from our distribution how disc lifetime would be affected by binarity. Assuming 20\% of stars have shorter disc lifetimes than average solely due to their binarity,
means that 20\% of stars would have a disc lifetime significantly shorter than 5 -- 10 Myr.
Thus, the \emph{median} disc lifetime of low-mass binary stars would be 1 Myr - 2 Myr shorter than those of single stars. However, we find it unlikely that a single cause is responsible for the shorter-than-average disc lifetimes.

The processes described in sections 5.1 -- 5.5 might influence a particular disc's lifetime. However, the relative importance of the different mechanisms for shaping the disc lifetime distribution still needs further exploration.

\section{Why the lifetime distribution matters}

Mean or median disc lifetimes have been used since \mbox{$>$20 years} to restrict the time available for planet formation. Nevertheless, the more than 20 Myr-old protoplanetary discs show that shifting towards a description by a disc lifetime distribution is long overdue. Such a disc lifetime distribution goes beyond acknowledging that a wide diversity exists in disc lifetimes. It quantifies the relative occurrence rate of very short and very long-lived discs. Proceeding to a disc lifetime effects mainly two research areas: 

First, disc dispersal theory must account for the spread in disc lifetimes. Several theoretical works have recently investigated how the disc lifetime depends on stellar mass \citep{Wilhelm:2022,Coleman:2022}. They agree with observations showing such a mass dependence. However, the question is whether the disc lifetime spread is primarily due to the stellar mass dependence. Our results show that the diversity of disc lifetimes remains when concentrating on just low-mass stars.  

Numerical investigations of disc dispersal often consider several of the many disc dispersal mechanisms. However, a match between simulation results and the here-derived disc lifetime distribution does not necessarily prove that the model is realistic. The problem is that there are many free parameters in these simulations. For example, the $\alpha$-viscosity, accretion rate, initial disc mass, and size. There is some risk that a match between the simulations and the distribution suffers from a posteriori treatment. Several combinations of these free parameters likely lead to a good match with the simulations. Therefore, ab initio simulations are more suitable for a comparison with the disc life distribution. These a priori models provide the chance for a deeper understanding of the underlying dispersal mechanism(s). They start out with a very limited number of assumptions, and the initial conditions are well-defined by observations. Therefore, these models require few free parameters. Simply selecting parameters that match the distribution should be avoided, as these a posteriori approaches remain on a purely heuristic level.

Second, many planet formation models assume a specific typical disc lifetime. In the past, often in the 2 -- 3 Myr range or at least $<$ 10 Myr. Discs existing for more than 10 Myr have attracted relatively little attention. Realising that at least the discs around low-mass stars persist on average longer (5 Myr -- 10 Myr) affects planet formation models. This change in perspective also applies to the width of the disc lifetime distribution. 

Discs with different lifetimes likely directly influence the resulting planets and planetary systems. For example, assuming the standard formation channel for a typical gas giant takes 8 Myr. Then, early disc dispersal would stop the gas accretion process prematurely. The result would be a smaller gas envelope and a core-to-gas compared to the standard scenario. This example illustrates how short disc lifetimes could affect the type of planet forming. By contrast, long disc lifetimes affect the number of planets forming, thus the planetary system structure. An explicit scenario is the following: In slowly forming planetary systems, the formation of one planet can induce spiral arms. In these spiral arms, the gas and dust density is enhanced, triggering the formation of additional planets can be triggered. Unfortunately, the dependence of planet formation on disc lifetimes is challenging to test observationally.

\section{Conclusion}

The discovery of protoplanetary discs older than 20 Myr implies a large diversity in disc lifetimes. Some discs exist for just a few Myrs, while others live for dozens of Myr. This finding implies that mean disc lifetimes provide limited information for disc development in general. Here, we introduced a new method for determining the disc lifetime distribution. Such a disc lifetime distribution can function as input for planet formation theory. In addition,  it provides a stringent test for existing disc dispersal theories.

We determine the disc lifetime distribution based on the disc frequencies of clusters of different ages. We concentrated on low-mass stars (spectral type M3.7--M6, \mbox{ $M_s \approx $ 0.1 -- 0.25 \MSun),} in nearby ($<$200 pc) clusters and associations. The reason is that disc lifetimes depend on stellar mass. We want to exclude this dependency \citep{Michel:2021, Pfalzner:2022}. Among the clusters of the sample, Upper Sco and UCL/LLC are especially relevant as their large number of stars provides a high statistical significance \citep{Luhman:2021}. We tested Gaussian, Gamma, Log-normal, Webull, Log-logistic and Beta distributions for their ability to provide a heuristic fit to the observational data.  

If all stars were initially surrounded by a disc, meaning $f$(t=0) = 100\%, the disc lifetime distribution could be best described by a Weibull distribution with $k$=1.78 and \mbox{$\lambda$ = 9.15.} This left-skewed distribution shows a median disc lifetime \mbox{$t_d$= 8.1 Myr,} with most discs dissipating at the age of \mbox{$t_{max}$ = 7.21 Myr.} However, we generally find that fits with initial disc fractions $<$100\% lead to a better match to the observational data. This finding agrees with the result of \citet{Michel:2021}. We obtain the best fit for an initial disc fraction of 65\%. A Weibull distribution with $k$=2.34 and $\lambda$ = 11.22 and a Gaussian distribution with $\sigma$=5.04 and $\mu$=9.52) both lead to similar good approximations for the disc lifetime distribution.
For the Weibull distribution, the corresponding median disc lifetime is 9.93 Myr and the maximum of the distribution is 8.83 Myr. 

This work is the first step in determining the disc lifetime distribution. One critical point is that the derived disc lifetime distribution is dominated by two clusters -- Upper Sco and UCL/LCC. Future improvements rely on more and better disc fraction data becoming available for clusters in the age range 3 Myr --  15 Myr. These data must be statistically significant, meaning each sample should contain $>$ 200 stars. A major challenge will be better constraining the disc fraction at cluster age t = 0.

Independent of the tested type, all distributions are extensive with $\sigma >$ 6 Myr) in each case. Understanding the underlying reason for this large width is crucial for a deeper understanding of the disc dispersal mechanism. The broad distribution of disc lifetimes is likely one of the main reasons for the diversity of planetary systems.

\acknowledgments
We thank the referee for their valuable suggestions on improving the manuscript.

\bibliographystyle{aa} 
\bibliography{references} 

\newpage
\appendix

\begin{table*}[h]
\caption{Disc fractions}
\centering \begin{tabular}{lrclllcclccccc}
\hline
Identification & $d$ & Age & $N_{stars}$  & $f_d$ & Limit  & Median mass &  log($\rho_c)$ &Source  \\
& pc  & Myr &              &           &    & [\MSun]& [\MSun/pc$^3$] \\
\hline
       Alessi 30         &  108 & 30       & 162  & 0.049\footnote{possibly debris disk fraction} &     0.04 \MSun & & & a)  \\
       UCL/LLC           &   150 & 15--20  & 3665 & 0.09$\pm$0.01            &   & 0.15                               &  -0.85 -(-1.05) & b) \\    
       32 Ori       &   95  & 15--20 &  160 & 0.07$^{+0.03}_{-0.02}$ &   & 0.15  & & g)\\
       Upp Sco           &  145 & 10--12  & 1774 & 0.22$\pm$0.02/0.20       & 0.01 \MSun  & 0.15                                &-0.59 &   b, c) \\
       Lupus-off cloud  & 160  & 10--12  &  60    & 0.21 $\pm$ 0.06 &  0.05 \MSun        &                                  & &   h)  \\
       $\eta$ Cha        &   94 &  8--14  &  40    & 0.28$\pm$ 0.14/33$\pm$0.16         &          &                                  & &   d)\\
       TW Hya            &   56 & 7--13   &  56   & 0.25/0.30, 0.19$^{+0.08}_{0.06}$       &          &                                   &  & d), j)\\
       Lupus-on cloud    &  160 & 6       &  30  & 0.63 $\pm$ 0.04 &   0.05 \MSun       &                                    & & h) \\
       CrA               &  152 & 5       & 146  & 0.23 $\pm$ 0.4  & 0.04 \MSun &                                   & & d)\\  
       $\epsilon$ Cha    &  101 & 5 (3--8)& 90& 0.5/0.3\footnote{The disk fraction is much higher in the center than the outskirts ($>$10 pc) of $\eta$ Cha}&      &                                                                                      & & d, e)\\
       Lupus             &  158 & 2.6--3.1 & 158  & 0.50/0.53  & 0.03 \MSun &                                      & & a, d) \\
       Cham I            &  188 & 1.7      & 183  & 0.44  & 6 $<$G $<$20    &                                      & & f) \\
       Cham II           &  197 & 1.7      & 41   & 0.76  & G12-G18         &                                      & & f)  \\
       Taurus            &  128-196 & 1--2 &  137    & 0.49, 0.637  &       0.05 \MSun          &                                   & & i, l)    \\
       Ophiuchus         &  139 & 1--2     &  420    & 0.62  &                 &                                   & & i) \\    
\hline  
\end{tabular}
\tablerefs{a) \citet[][]{Galli:2021_1}, b) \citet[][]{Luhman:2021}, c) \citet{Luhman:2020}, d) \citet[][]{Michel:2021}, e) \citet[][]{Dickson:2021}, f) \citet[][]{Galli:2021_2},  g) \citep{Luhman:2022}, h) \citep{Luhman:2020}, \i) \citet[][]{Manzo:2020},  j) \citep{Luhman:2023}, \mbox{l) \citep{{Luhman:2023b}}.}
              }
\label{tab:cluster_properties}
\end{table*}

\begin{table*}[h]
        \caption{Fit parameters for the various distributions}
        \centering\begin{tabular}{lrcllccclcccccc}
        \hline
        \multicolumn{1}{l}{ID}    & \multicolumn{2}{l}{Disc fractions} & \multicolumn{2}{c}{Cluster ages} &
         \multicolumn{2}{l}{Gamma} &  \multicolumn{2}{l}{Beta} &  \multicolumn{2}{l}{Log-normal}  &  \multicolumn{2}{l}{Weibull} &  \multicolumn{2}{l}{Log-logistic} \\ 
        ID & $f_{USco}$ & $f_{U/L}$ & $t_{USco}$ & $t_{U/L} $& $\bar{t_d}$ & $\sigma(t_d)$  & $\bar{t_d}$ & $\sigma(t_d)$ & $\bar{t_d}$ & $\sigma(t_d)$ & $\bar{t_d}$ & $\sigma(t_d)$ & $\bar{t_d}$ & $\sigma(t_d)$ \\ 
        \hline
        L1$_{100}$  & 0.25 & 0.09  & 11.0 & 15.0 & 8.5  & 20.6 & 8.4 & 20.8 & 8.9 & 19.4 & 8.1 & 22.4 & 9.3 & 24.8\\
        L2$_{100}$  & 0.22 & 0.09  & 11.0 & 17.0 & 7.3  &  46.7 & - & - & 8.2 & 53.1 & 7.3 & 46.6 & 8.8 & 140.3\\
        L1$_{80}$  & 0.25 & 0.09  & 11.0 & 15.0 &  9.4  &  19.9 & 9.3 & 20.1 & 9.8 & 19.2 & 9.0 & 21.7 & 10.0 & 25.5\\
        L2$_{80}$  & 0.22 & 0.09  & 11.0 & 17.0 &  8.5  &  48.7 & 8.3 & 47.0 & 9.2 & 55.6 & 8.4 & 48.3 & 9.8 & 135.2\\
        L3$_{100}$  & 0.25 & 0.09  & 12.0 & 15.0 & 9.8  & 13.3 & 9.8 & 13.6 & 10.1 & 12.2 & 9.3 & 16.2 & 10.4 & 13.8\\
        L4$_{100}$  & 0.22 & 0.11  & 11.0 & 15.0 & 7.5  &  36.8 & 7.3 & 36.0 & 8.2 & 39.9 & 7.3 & 36.8 & 8.7 & 80.8\\
 \hline
 \hline      
              \end{tabular}
        \label{tab:fit_parameters_Gauss}
\end{table*}

\begin{table}[h]
        \caption{Deviations of different distributions}
        \centering\begin{tabular}{llllccclccccc}
        \hline
Type & $D$(100\%) & $D$ (80\%) &  $D$ (65\%) \\
        \hline
Gaussian   & 0.0564 & 0.0424 & 0.0302   \\
Gamma      & 0.0828 & 0.0558 &  0.0603 \\
Log-normal & 0.0873 & 0.0579 &  0.0368\\
Weibull    & 0.0769 & 0.0523 &  0.0340 \\
Log-logistic &  0.0886 & 0.0582 & 0.0356\\
Beta       & 0.0815  & 0.0552 & 0.0356\\
        \hline
 \hline
              \end{tabular}
        \label{tab:D-values}
        \begin{flushleft}
        \tablecomments{Column~1 indicates the distribution type,  columns 2 -- 4 show the $D$ values corresponding to the distributions listed in table \ref{tab:fit_distributions}.  }
        \end{flushleft}
\end{table}

\end{document}